\documentclass[a4paper,11pt]{article}
\pdfoutput=1 

\usepackage{jheppub} 
\usepackage{bbm,bm,graphicx,mathtools,color,slashed,hyperref}
\usepackage{amssymb,amsmath}
\usepackage{dsfont}
\usepackage{braket}

\usepackage{subfig}

\newcommand{\diff}{\mathrm{d}}

\newcommand{\tr}{\operatorname{tr}}

\newcommand{\im}{\mathrm{i}}
\newcommand{\calA}{\mathcal{A}}

\newcommand{\calR}{\mathcal{R}}
\newcommand{\calP}{\mathcal{P}}

\newcommand{\rmd}{\mathrm{d}}
\newcommand{\rme}{\mathrm{e}}

\title{
Center-vortex semiclassics with non-minimal 't~Hooft fluxes on $\mathbb{R}^2\times T^2$ and center stabilization at large $N$
}

\author[1]{Yui Hayashi,}
\affiliation[1]{Yukawa Institute for Theoretical Physics,
Kyoto University, Kyoto, 606-8502, Japan}
\emailAdd{yui.hayashi@yukawa.kyoto-u.ac.jp}

\author[1]{ Yuya Tanizaki,}
\emailAdd{yuya.tanizaki@yukawa.kyoto-u.ac.jp}

\author[2]{ Mithat \"{U}nsal}
\affiliation[2]{Department of Physics, North Carolina State University, Raleigh, NC 27607, USA}
\emailAdd{unsal.mithat@gmail.com}

\abstract{
We consider the semiclassical description of confinement for $4$d $SU(N)$ Yang-Mills theory on small $\mathbb{R}^2\times T^2$ with non-minimal 't~Hooft twist $p$ with $\gcd(N,p)=1$. 
For this purpose, we construct the self-dual center vortex for non-minimal 't~Hooft twists from the Kraan-van~Baal-Lee-Lu-Yi (KvBLLY) monopoles by using the $3$d Abelianized description of $SU(N)$ gauge fields on $\mathbb{R}^3\times S^1$ with nontrivial holonomy backgrounds. 
This construction shows the self-dual vortex has (1) the fractional magnetic charge $q/N$ with $pq=1$ mod $N$, (2) the fractional topological charge $1/N$, and (3) the fractional instanton action $S_{\mathrm{YM}}=8\pi^2/(Ng^2)$. 
The confinement vacua for $NL\Lambda\ll 1$ can be described by the dilute gas approximation of center vortices, and we give the semiclassical formula for the $\theta$ dependence and confining string tensions. 
We apply this result to understand the suitable choice of the twist $p$ for center stabilization at large $N$. In particular, we test the proposal using the Fibonacci sequence, $N=F_{n+2}$ and $p=F_n$, suggested in studies of the twisted Eguchi-Kawai model, from the viewpoint of the $1$-form and $0$-form center symmetries.  
}

\preprint{YITP-25-57}

\begin{document}

\maketitle

\section{Introduction and summary}
\label{sec:introduction}

$4$-dimensional gauge theories provide important playgrounds of strongly coupled systems due to asymptotic freedom, and they produce a lot of interesting physical phenomena, such as quark confinement~\cite{Wilson:1974sk}, chiral symmetry breaking~\cite{Nambu:1961tp, Nambu:1961fr}, fractional $\theta$ dependence, etc. 
To understand the permanent quark confinement, 't~Hooft proposed to consider pure $SU(N)$ Yang-Mills (YM) theory inside a box with a twisted boundary condition~\cite{tHooft:1979rtg, tHooft:1981sps}, which we call 't~Hooft twist or 't~Hooft flux in an interchangeable way in this paper.  

The 't~Hooft twist corresponds to the twisted boundary condition associated with the center symmetry: The center symmetry is nowadays understood in the language of the $1$-form symmetry~\cite{Gaiotto:2014kfa}, and we can introduce its background $2$-form gauge field~\cite{Kapustin:2013uxa, Kapustin:2014gua}, which is nothing but the 't~Hooft flux. 
Moreover, there was a discovery of new 't~Hooft anomalies associated with the $1$-form symmetry~\cite{Gaiotto:2017yup}, constraining the nonperturbative dynamics of YM theory via anomaly matching condition, and it is crucial to introduce the background $2$-form gauge field to detect those new anomalies.
It brings the renewed interest of studying gauge theories inside a box with the 't~Hooft flux background~\cite{Yamazaki:2017ulc, Cox:2021vsa, Tanizaki:2022ngt, Poppitz:2022rxv} since the nontrivial 't~Hooft flux plays the key role to maintain the anomaly matching constraint after dimensional reduction~\cite{Shimizu:2017asf, Tanizaki:2017qhf, Yamazaki:2017dra, Tanizaki:2017mtm, Dunne:2018hog, Yonekura:2019vyz}. 

In a series of works~\cite{Tanizaki:2022ngt, Tanizaki:2022plm, Hayashi:2023wwi, Hayashi:2024qkm, Hayashi:2024gxv, Hayashi:2024yjc, Hayashi:2024psa, Guvendik:2024umd}, the present authors have developed the semiclassical description of confinement on small $\mathbb{R}^2\times T^2$ with the minimal 't~Hooft twist $n_{34}=1$ without suffering from infrared divergence: 
The classical vacuum with the 't~Hooft twist preserves the $\mathbb{Z}_N\times \mathbb{Z}_N$ center symmetry and introduces the perturbative mass gap by $\frac{2\pi}{NL}$ for the $2$d effective theory, which ensures the validity of the weak-coupling analysis if $NL\Lambda\ll 1$. 
To describe the area law of the Wilson loop in $\mathbb{R}^2$, we need to take into account the nonperturbative effect described by the center vortex, which carries the fractional magnetic flux to rotate the phase of Wilson loops and also has the $\frac{1}{N}$ topological charge. 
Then, the confinement vacua for $NL\Lambda\ll 1$ turn out to be described by the dilute gas of center vortices, and this semiclassical theory successfully reproduces the multi-branch structure of the $\theta$ vacua~\cite{Witten:1980sp, DiVecchia:1980yfw, tHooft:1981bkw, Witten:1998uka} of the $4$d $SU(N)$ YM theory. 
Moreover, this framework gives an explicit derivation of the chiral effective Lagrangian including the anomalous $U(1)$ axial contribution for various representations of Dirac fermions~\cite{Tanizaki:2022ngt, Tanizaki:2022plm, Hayashi:2023wwi, Hayashi:2024qkm, Hayashi:2024gxv}.

It is then natural to ask if this weak-coupling confinement for $NL\Lambda\ll 1$ is adiabatically connected to the strongly-coupled confinement on $\mathbb{R}^4$ through decompactification without having phase transitions. 
For the case of $SU(2)$, Refs.~\cite{Soler:2025vwc, Bergner:2025qsm} recently reported promising results of the lattice numerical simulations that suggest smooth crossover between the weakly-coupled gas of center vortices and the strongly-coupled confinement state. 
Thus, we are interested in whether the center stabilization is achievable in the other limit, $N \gg 1$.  

To get a hint for addressing this question, let us explain the related problem of the twisted Eguchi-Kawai (TEK) 1-site model~\cite{GonzalezArroyo:1982hz, GonzalezArroyo:1982ub, GonzalezArroyo:2010ss}. 
Eguchi and Kawai~\cite{Eguchi:1982nm} pointed out that large-$N$ gauge theories show volume independence for single-trace observables as long as the center symmetry is preserved, and thus properties of confinement can be described by the $1$-site large-$N$ matrix model in principle. 
The naive 1-site model in the original proposal spontaneously breaks the center symmetry in the large-$N$ limit, and Gonzalez-Arroyo and Okawa~\cite{GonzalezArroyo:1982hz, GonzalezArroyo:1982ub} proposed the TEK model that introduces the 't~Hooft twists to stabilize the center-symmetric vacuum.\footnote{
For our $\mathbb{R}^2\times T^2$ setup, we introduce the twist only along the $3$-$4$ direction, $n_{34}=p$, and we require $\gcd(N,p)=1$ to preserve the $\mathbb{Z}_N\times \mathbb{Z}_N$ center symmetry completely. In the context of TEK, however, all the directions are compactified, which makes it impossible to preserve $(\mathbb{Z}_N)^4$ completely. To preserve maximal center symmetries and the Euclidean hypercubic rotation, the possible choice is to take $N=M^2$ with some integer $M$ and $n_{\mu\nu}=k M$ for all $\mu<\nu$ with some $k$ that is coprime with $M$: Twist eaters classically preserve $(\mathbb{Z}_M)^4\subset (\mathbb{Z}_N)^4$ for this symmetric twist. Thus, the relation between $p$ and $N$ in our context has to be translated to the one between $k$ and $M$ for the TEK model. } 
While the 't~Hooft twist classically prefers the center-symmetric vacuum, fluctuations induce tachyonic instabilities for the TEK model with the fixed twists at some intermediate gauge couplings~\cite{Guralnik:2002ru, Bietenholz:2006cz, Teper:2006sp, Azeyanagi:2007su}. 
There are two approaches to evade this tachyonic instability: One is to make the 't~Hooft twist judiciously dependent on $N$~\cite{GonzalezArroyo:2010ss}, and the other is to introduce massive adjoint fermions~\cite{Azeyanagi:2010ne}. 
For the first option, the appropriate choice for the twist has been investigated in detail~\cite{GonzalezArroyo:2010ss, Gonzalez-Arroyo:2014dua, Perez:2017jyq, Chamizo:2016msz, GarciaPerez:2018fkj, Bribian:2019ybc}, and the TEK model is providing a powerful tool for numerically studying large-$N$ gauge theories~\cite{Gonzalez-Arroyo:2012euf, GarciaPerez:2014azn, Gonzalez-Arroyo:2015bya, Perez:2020vbn, Bonanno:2023ypf, Bonanno:2024bqg, Bonanno:2024onr}. 
Related to the second option, there is a fruitful development of weak-coupling description of confinement on $\mathbb{R}^3\times S^1$~\cite{Davies:1999uw, Davies:2000nw, Unsal:2007vu, Unsal:2007jx, Unsal:2008ch, Shifman:2008ja, Poppitz:2008hr, Poppitz:2012sw}.

Tachyonic instability appears also for the case of our $\mathbb{R}^2\times T^2$ setup with the minimal 't~Hooft twist for $N\gg 1$~\cite{Guralnik:2002ru,  Bietenholz:2006cz, Guvendik:2024yzh}, and 
the observation of the TEK model motivates us to study the center-vortex semiclassics with non-minimal 't~Hooft twists, $n_{34}=p$. 
While it might sound as if it is straightforward to extend our previous work~\cite{Tanizaki:2022ngt} from minimal to non-minimal twists, it is not the case as we need to answer the following question: What is the center vortex on $\mathbb{R}^2\times T^2$ with the non-minimal twist?

To identify the properties of the center vortex that dominantly contributes to the semiclassics, we have to solve the self-dual YM equation on $\mathbb{R}^2\times T^2$ with the 't~Hooft twisted boundary condition. 
However, its analytic knowledge with 't~Hooft twists is quite limited.\footnote{
There exists a constant field-strength solution by 't~Hooft, but it solves the self-duality equation only with a specific aspect ratio of $T^4$~\cite{tHooft:1981nnx}. 
Deforming this constant field-strength solution with the detuning parameter on the torus sizes is recently investigated extensively by Anber and Poppitz in Refs.~\cite{Anber:2022qsz, Anber:2023sjn, Anber:2024mco, Anber:2025yub}. 
For analytic results on $\mathbb{R}^2\times T^2$, Ford and Pawlowski, in Refs.~\cite{Ford:2002pa, Ford:2003vi, Ford:2005sq}, construct the $\frac{1}{2}$-instanton of $SU(2)$ by cutting the doubly-periodic instanton with a specific moduli parameter into halves.} 
Thus, most studies of the self-dual YM equation with 't~Hooft twist rely on numerical minimization of the lattice Wilson action: 
The fractional instanton on $\mathbb{R}\times T^3$ is found in Refs.~\cite{GarciaPerez:1989gt, GarciaPerez:1992fj, Itou:2018wkm, DasilvaGolan:2022jlm, Wandler:2024hsq}, 
and the one on $\mathbb{R}^2\times T^2$ is found in Refs.~\cite{Gonzalez-Arroyo:1998hjb, Montero:1999by, Montero:2000pb}. 
When developing the semiclassics on $\mathbb{R}^2\times T^2$ with the minimal 't~Hooft twist in Ref.~\cite{Tanizaki:2022ngt}, the authors relied on the numerical results of Refs.~\cite{Gonzalez-Arroyo:1998hjb, Montero:1999by, Montero:2000pb} to identify the properties of the center vortex. 
To understand the semiclassics with non-minimal twists, it would be evident that we need a more systematic way to understand the properties of the center vortex.

Such a method recently becomes available thanks to the discovery of Ref.~\cite{Hayashi:2024yjc} that uncovers the connection between the center vortex on $\mathbb{R}^2\times T^2$ with the 't~Hooft twist and the monopole constituents of the $4$d instanton on $\mathbb{R}^3\times S^1$~\cite{Lee:1997vp, Lee:1998bb, Lee:1998vu, Kraan:1998kp, Kraan:1998pm, Kraan:1998sn}. 
We call those monopole constituents as Kraan-van~Baal-Lee-Lu-Yi (KvBLLY) monopoles. 
When introducing the hierarchy between the sizes of $T^2$ as $L_3\gg N L_4$ with a nontrivial holonomy along $(S^1)_{L_4}$, we can use the Abelianized description for the $SU(N)$ gauge field on $\mathbb{R}^2\times (S^1)_{L_3}$, and the 't~Hooft twist on $T^2$ becomes the center-twisted boundary condition along $(S^1)_{L_3}$.  
When we put the KvBLLY monopole on $\mathbb{R}^2\times (S^1)_{L_3}$, the magnetic flux emitted from monopoles forms a tube that wraps around the $(S^1)_{L_3}$ direction due to the center-twisted boundary condition, and this is exactly the center vortex that solves the self-dual equation on $\mathbb{R}^2\times T^2$~\cite{Hayashi:2024yjc, Hayashi:2024psa, Guvendik:2024umd}.\footnote{
If we restrict our attention to the $SU(2)$ case, this idea turns out to be essentially equivalent with the construction of the $\frac{1}{2}$-instanton from the doubly-periodic instanton in Refs.~\cite{Ford:2002pa, Ford:2003vi, Ford:2005sq}. } 
This knowledge enables us to develop the center-vortex semiclassics with non-minimal twists, and we study its properties for general non-minimal twist $p$ with $\gcd(N,p)=1$ in Section~\ref{sec:nonminimal_twist}. 
We derive the semiclassical formula for the $\theta$-dependent vacuum structure and confining string tensions using dilute center-vortex gas, and we also discuss the semiclassical realization of generalized anomaly matching in detail.

We here used the Abelianized description of $SU(N)$ gauge fields on $\mathbb{R}^3\times S^1$ to solve the self-dual YM equation on $\mathbb{R}^2\times (T^2)_{p/N \text{ flux}}$ in the asymmetric limit, but it also offers the framework to develop the semiclassics for weakly-coupled confinement on $\mathbb{R}^3\times S^1$ with adjoint fermions~\cite{Unsal:2007vu, Unsal:2007jx, Unsal:2008ch}. 
The semiclassical confinement on $\mathbb{R}^3\times S^1$ is caused by the dual Debye screening via Coulomb gas of KvBLLY monopoles (and/or magnetic bions), where the long-range interaction between monopole-constituents of $4$d instanton plays the pivotal role. 
It may seem to have a sharp contrast with the $2$d center vortex that has short-range interaction, but, remarkably, we here observe the explicit transmutation between monopole-constituents and vortex-constituents of $4$d instanton thanks to the 't~Hooft twist~\cite{Hayashi:2024yjc, Hayashi:2024psa, Guvendik:2024umd}. 
We cannot help but imagine that there must be a deeper structure inside $4$d instanton than meets the eye, which encourages us to examine it further in future studies. 

\begin{figure}[t]
\centering
\includegraphics[angle=0, width=0.99\textwidth]{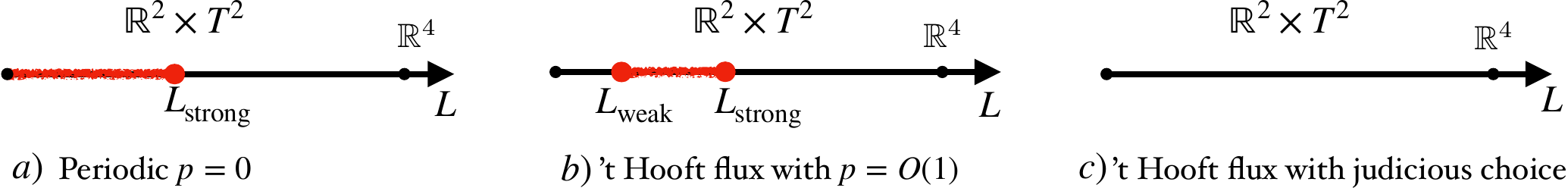}
\caption{Boundary-condition dependence for the phase structures of $4$d $SU(N)$ YM theory on $\mathbb{R}^2\times T^2$ with the torus size $L$ for $N\gg 1$. Confinement regions are shown with black straight lines, and center-broken regions are drawn with red crayon lines.\\
\textbf{\textit{a})}~Periodic compactification, $p=0$, has a phase transition around the strong scale, $L_{\text{strong}} \sim \Lambda^{-1}$. For the periodic case, the weak-coupling regimes are always deconfined.\\
\textbf{\textit{b})}~With an 't Hooft flux $p\sim O(1)$, weakly-coupled and strongly-coupled confinement states are separated by intermediate center-broken states if $N\gg 1$. 
Upper boundary is strongly coupled, $L_{\mathrm{strong}}\sim \Lambda^{-1}$, and the lower boundary is characterized as  $L_{\text{weak}} \sim \mathrm{e}^{-O(N^2)} \Lambda^{-1}$.\\
\textbf{\textit{c})}~Adiabatic continuity is achievable for judiciously chosen pairs of $(N,p)$, e.g. $N=F_{n+2}$ and $p=F_n$, even for $N\gg 1$. Weakly coupled center-vortex gas is continuously connected to the strongly coupled confining phase, which is the main target of this paper. 
}
\label{fig:phases}
\end{figure}

We then consider the application of our semiclassical result to investigate the large-$N$ center stabilization. 
In Section~\ref{sec:CenterStabilization}, we shall discuss the center-stabilizing criterion using the semiclassical analysis. 
It turns out that Wilson loops with $p$-box representations are deconfined in the large-$N$ limit, and it suggests that any $O(1)$ multiples of $p$ should not become $O(1)$ in mod $N$ in order not to spontaneously break the $1$-form symmetry as $N\to \infty$. 
The center-stabilizing criterion we obtained is identical to the one found in the studies of $3$d YM theory on $\mathbb{R}\times T^2$ to maintain the $0$-form center symmetry in Refs.~\cite{Chamizo:2016msz, GarciaPerez:2018fkj, Bribian:2019ybc}, 
and those studies uncovered the choice $(N,p)=(F_{n+2},F_n)$ with the Fibonacci sequence is particularly good for $0$-form center stabilization. 
We test this proposal for the large-$N$ string tensions and find the consistency of the large-$N$ center stability.  
Our computation of large-$N$ string tensions provides a nontrivial test of the proposal, showing that one can simultaneously achieve the large-$N$ center stabilization for both the $0$-form and $1$-form center symmetries. 
Adiabatic continuity between the weakly-coupled and strongly-coupled confinement states for $N\gg 1$ is realized with a suitable choice of $p$ as shown in Figure~\ref{fig:phases}.

In Appendix~\ref{sec:ActionvsEntropy}, we give the action-versus-entropy argument to obtain the $0$-form center-stabilizing criterion in our $\mathbb{R}^2\times T^2$ setup after replacing $T^2$ by the $1\times 1$ lattice. 
Appendix~\ref{sec:ContinuedFraction} summarizes useful properties of the continued fraction, and Appendix~\ref{sec:RootWeight} gives the convention for the $SU(N)$ root and weight vectors. 

\section{Semiclassics on \texorpdfstring{$\mathbb{R}^2\times T^2$}{R2xT2} with non-minimal 't~Hooft flux}
\label{sec:nonminimal_twist}

In this section, we extend the previous study~\cite{Tanizaki:2022ngt} by considering the $4$d $SU(N)$ YM theory on $\mathbb{R}^2\times T^2$ with a non-minimal 't~Hooft flux $n_{34}=p$ with $\gcd(p,N)=1$. 
To develop the semiclassical framework, the important task is to identify the center-vortex configuration with the minimal classical action. We construct such a center vortex from the KvBLLY monopoles by applying the observation in Refs.~\cite{Hayashi:2024yjc, Hayashi:2024psa}, and we calculate the vacuum structure and string tensions using the dilute gas approximation of center vortices. 

\subsection{Classical analysis with non-minimal flux}
\label{sec:classical}

The first step to develop the semiclassical description is to understand the classical vacuum of $4$d $SU(N)$ YM theory on $\mathbb{R}^2\times T^2 \ni (\bm{x},x_3,x_4)$ with the 't~Hooft twisted boundary condition and also its perturbative spectrum. 
Here, $x_3+L_3\sim x_3$ and $x_4+L_4\sim x_4$. 

To find the classical vacuum with the 't~Hooft twisted boundary condition, $n_{34}=p$, it is convenient to consider the lattice regularized Wilson action. We require $p$ is coprime with $N$, i.e. $\gcd(N,p)=1$. The effect of the 't~Hooft twisted boundary condition in the $3$-$4$ direction can be translated into the center twist of the lattice coupling constant as 
\begin{equation}
    S_{\mathrm{W}}=\frac{1}{g^2}\sum_x\sum_{\mu\not=\nu} \tr\left(\bm{1}_{N\times N}-\rme^{-\frac{2\pi\im}{N} B_{\mu\nu}(x)}U_\mu(x)U_\nu(x+\hat{\mu})U^\dagger_{\mu}(x+\hat{\nu})U^\dagger_\nu(x)\right), 
    \label{eq:WilsonActon_ZN2form}
\end{equation}
where $B_{\mu\nu}(x)=-B_{\nu\mu}(x)$ for $\mu<\nu$ is given by
\begin{equation}
    B_{\mu\nu}(x)=\left\{
    \begin{array}{cl}
        p  & \quad ((\mu,\nu)=(3,4),\, (x_3,x_4)=(L_3-1,L_4-1)), \\
        0 & \quad (\text{otherwise}). 
    \end{array}
    \right.
\end{equation}
This plaquette field $B_{\mu\nu}(x)$ can be understood as the $\mathbb{Z}_N$-valued $2$-cochain, and it is related to the 't~Hooft twist by $\int_{(T^2)_{\mu\nu}}B=n_{\mu\nu}$. 
For details, see Appendix~A of Ref.~\cite{Tanizaki:2022ngt}. 
To minimize this action along the $3$-$4$ direction, let us introduce the $SU(N)$ clock and shift matrices by 
\begin{equation}
    C=\omega^{-\frac{N-1}{2}}
    \begin{pmatrix}
        1&      &      &    \\
         &\omega&      &    \\
         &      &\ddots&    \\
         &      &      & \omega^{N-1}
    \end{pmatrix}, \quad
    S=\omega^{-\frac{N-1}{2}}
    \begin{pmatrix}
        0& 1 &      &   \\
         & 0 & \ddots&   \\
         &   &\ddots& 1  \\
        1&   &      & 0
    \end{pmatrix}, 
    \label{eq:ShiftClockMatrix}
\end{equation}
where $\omega=\rme^{2\pi \im/N}$ is the $N$-th root of unity, and they satisfy 
\begin{equation}
    SC=\omega\, CS.  
    \label{eq:ShiftCloclAlgebra}
\end{equation}
If we can find a set of link variables that satisfy
\begin{equation}
    \rme^{-\frac{2\pi \im}{N} B_{34}(x)}U_3(x)U_4(x+\hat{3})U^\dagger_{3}(x+\hat{4})U^\dagger_4(x) = \bm{1}_{N\times N}, 
\end{equation}
then it obviously minimizes the action along the $3$-$4$ direction. 
This sufficient condition can be actually achieved by setting 
\begin{equation}
    U_3(\bm{x},L_3-1,x_4)=S^{-p},\quad U_4(\bm{x},x_3,L_4-1)=C^{-1}, 
\end{equation}
and $U_{\mu=3,4}(x)=\bm{1}_N$ for other links. We can readily check this is the unique solution up to gauge transformations by taking the complete axial gauge at $(x_3,x_4)=(0,0)$. 

As a corollary of the uniqueness, the $\mathbb{Z}_N\times \mathbb{Z}_N$ center symmetry is unbroken at the classical vacuum:\footnote{This causes the huge difference compared with the case of periodic $T^2$ or $S^1$ compactified setups. In those cases, the holonomy is a classical moduli, and the Gross-Pisarski-Yaffe one-loop potential~\cite{Gross:1980br} gives the leading contribution to discuss the realization of the $0$-form center symmetry. 
In our twisted $T^2$ case, however, the classical action gives the leading contribution and prefers the center-symmetric vacuum. } The holonomies of the $3$ and $4$ directions are given by
\begin{equation}
    P_3=S^{-p},\quad P_4=C^{-1}, 
    \label{eq:ClassicalVacuum}
\end{equation}
and it satisfies $\tr(P_3^{n_3}P_4^{n_4})=0$ unless both $n_3, n_4$ are multiples of $N$. 

Let us now go back to the continuum description, and we discuss the perturbative spectrum around this center-symmetric classical vacuum. 
In the continuum, we introduce the $SU(N)$-valued transition functions, $g_3(\bm{x},x_4)$ and $g_4(\bm{x},x_3)$, when identifying $x_{3,4}+L_{3,4}\sim x_{3,4}$ for $T^2$, and the 't~Hooft twist imposes $g_3(\bm{x},L_4)g_4(\bm{x},0)=\omega^p g_4(\bm{x},L_3)g_3(\bm{x},0)$. 
We may choose a gauge so that $g_3(\bm{x},x_4)=S^{-p}$ and $g_4(\bm{x},x_3)=C^{-1}$, and then the gauge field $a_\mu(x)$ obeys the following boundary condition,
\begin{align}
    a_\mu(\bm{x},x_3+L_3,x_4)&=S^{p} a_\mu(\bm{x},x_3,x_4) S^{-p}, \notag\\
    a_\mu(\bm{x},x_3,x_4+L_4)&=C\, a_\mu (\bm{x},x_3,x_4) C^{-1}.
\end{align}
The classical vacuum is given by $a_\mu=0$, which reproduces \eqref{eq:ClassicalVacuum}. 
For the Fourier expansion of this twisted boundary condition, the following basis of the $\mathfrak{su}(N)$ algebra~\cite{GonzalezArroyo:1982hz} is useful, 
\begin{equation}
    J_{(\ell_3,\ell_4)}=\omega^{-\frac{q}{2}\ell_3\ell_4}C^{-q\ell_3}S^{\ell_4},
    \label{eq:SUN_clockshiftbasis}
\end{equation}
where we introduce another integer $q$ by
\begin{equation}
    p\, q=1 \bmod N. 
\end{equation}
We note such a choice of $q$ exists due to $\gcd(N,p)=1$ and it is unique in mod $N$. 
Since $J_{(\ell_3,\ell_4)}^\dagger = J_{(-\ell_3,-\ell_4)}$, $a_\mu$ can be expanded as 
\begin{equation}
    a_\mu(\bm{x},x_3,x_4)=\sum_{(\ell_3,\ell_4)\not=0}a_\mu^{(\ell_3,\ell_4)}(\bm{x},x_3,x_4)J_{(\ell_3,\ell_4)}, 
\end{equation}
with $(a_\mu^{(\ell_3,\ell_4)})^*=a_\mu^{(-\ell_3,-\ell_4)}$. In this basis, the twisted boundary condition is simplified as 
\begin{align}
    a_\mu^{(\ell_3,\ell_4)}(\bm{x},x_3+L_3,x_4)&= \omega^{-\ell_3}a_\mu^{(\ell_3,\ell_4)}(\bm{x},x_3,x_4), \notag\\ 
    a_\mu^{(\ell_3,\ell_4)}(\bm{x},x_3,x_4+L_4)&= \omega^{-\ell_4}a_\mu^{(\ell_3,\ell_4)}(\bm{x},x_3,x_4), 
\end{align}
and the Fourier expansion is given by 
\begin{equation}
    a_\mu^{(\ell_3,\ell_4)}(\bm{x},x_3,x_4)=\sum_{k_3,k_4\in \mathbb{Z}}\tilde{a}_\mu^{(Nk_3+\ell_3, Nk_4+\ell_4)}(\bm{x})\, \rme^{-\frac{2\pi\im}{N}\left((Nk_3+\ell_3)\frac{x_3}{L_3}+(Nk_4+\ell_4)\frac{x_4}{L_4}\right)}. 
\end{equation}
Therefore, the gauge field acquires the perturbative mass gap by 
\begin{equation}
    m_{(Nk_3+\ell_3, Nk_4+\ell_4)}^2=\left(\frac{2\pi}{N}\right)^2\left\{\frac{(Nk_3+\ell_3)^2}{L_3^2}+\frac{(Nk_4+\ell_4)^2}{L_4^2}\right\}. 
    \label{eq:PerturbativeGap}
\end{equation}
In particular, as $(\ell_3,\ell_4)\not = (0,0)$, the $2$d zero mode does not exist at all for this classical vacuum under the 't~Hooft twisted boundary condition. 

Due to the presence of the perturbative mass gap, we can integrate out all the particle-like excitations while keeping smallness of the gauge couplings if we take
\begin{equation}
    \frac{2\pi}{NL_3}, \frac{2\pi}{NL_4} \gg \Lambda, 
\end{equation}
where $\Lambda$ is the strong scale. Therefore, one can develop a semiclassical description of the confinement phenomenon in a reliable manner in this regime~\cite{Tanizaki:2022ngt}. 
Strictly speaking, this argument contains a subtle issue of the tachyonic instability in the large-$N$ limit, and we shall revisit this problem in Section~\ref{sec:0FormCenterStabilization} and also in Appendix~\ref{sec:ActionvsEntropy}. 

\subsection{Minimal-action center vortex from KvBLLY monopoles}
\label{sec:CenterVortexFromKvBLLY}

Under the $T^2$ compactification, the $4$d $\mathbb{Z}_N$ $1$-form symmetry, $(\mathbb{Z}_N^{[1]})_{4\rmd}$, splits as 
\begin{equation}
    (\mathbb{Z}_N^{[1]})_{4\rmd}\rightarrow \mathbb{Z}_N^{[1]}\times \mathbb{Z}_N^{[0]}\times \mathbb{Z}_N^{[0]}, 
    \label{eq:1FormSymmetry_compactification}
\end{equation}
where the $2$d $\mathbb{Z}_N^{[1]}$ symmetry acts on the Wilson loops living on the uncompactified $\mathbb{R}^2$ direction and the $0$-form $\mathbb{Z}_N$ symmetries act on $P_3$ and $P_4$. 
As we have seen in \eqref{eq:ClassicalVacuum}, $\mathbb{Z}_N^{[0]}\times \mathbb{Z}_N^{[0]}$ is unbroken at the classical vacuum. 
However, the $2$d $\mathbb{Z}_N^{[1]}$ symmetry is spontaneously broken at the classical level because of the Higgsing, which causes the topological degeneracy of the classical vacuum: If we further compactify $\mathbb{R}^2$ to $\mathbb{R}\times S^1$, then the classical value of $P_2$ must commute with both $P_3$ and $P_4$, and we obtain $N$-fold degenerate ground states, $P_2=\rme^{\frac{2\pi\im}{N}k}\bm{1}$, with $k=0,1,\ldots, N-1$. 

When a local field theory has degenerate classical ground states, there exists a soliton connecting them. 
If the degeneracy is caused by spontaneous breaking of $0$-form symmetries, such a soliton is a domain wall, i.e. a codim-$1$ excitation. 
In our case, as the classically broken symmetry is a $1$-form symmetry, the soliton is a vortex-like codim-$2$ object. 
By definition, this vortex rotates the phase of the Wilson loops on $\mathbb{R}^2$ by a center element, which is nothing but the center vortex~\cite{tHooft:1977nqb, Cornwall:1979hz, Nielsen:1979xu, Ambjorn:1980ms, DelDebbio:1996lih,  Faber:1997rp, DelDebbio:1998luz, Langfeld:1998cz, Kovacs:1998xm, Engelhardt:1999fd, deForcrand:1999our}. 

The center vortices on $\mathbb{R}^2\times T^2$ with the 't~Hooft flux carry fractional topological charges. 
Let us consider the center vortex which rotates the phase of the Wilson loop as 
\begin{equation}
    W(C)\rightarrow \rme^{\frac{2\pi\im}{N}q}W(C), 
\end{equation}
then we can single it out by compactifying $\mathbb{R}^2$ with introducing the 't~Hooft flux $-q$, 
\begin{equation}
    \mathbb{R}^2\times \underbrace{T^2}_{\text{$p/N$ flux}}\Rightarrow \underbrace{T^2}_{\text{$-q/N$ flux}}\times \underbrace{T^2}_{\text{$p/N$ flux}}. 
\end{equation}
As a result, the topological charge of the center vortex has to be~\cite{vanBaal:1982ag} 
\begin{equation}
    Q_{\mathrm{top}}\in \frac{pq}{N}+\mathbb{Z}. 
\end{equation}
The Bogomolnyi bound of the YM action is given by $S_{\mathrm{YM}}\ge \frac{8\pi^2}{g^2}|Q_{\mathrm{top}}|$. 
If we think that the Bogomolnyi bound can be saturated, there would exist a center (anti-)vortex that carries 
\begin{itemize}
    \item the fractional magnetic flux $\pm \frac{q}{N}$ with $pq=1$ mod $N$, 
    \item the fractional topological charge $Q_{\mathrm{top}}=\pm \frac{1}{N}$, and
    \item the fractional instanton action $S_{\mathrm{YM}}=\frac{8\pi^2}{Ng^2}$. 
\end{itemize}
In the following, we argue that such a center vortex exists and, 
moreover, it can be constructed out of the KvBLLY monopoles by introducing another hierarchy $L_3\gg NL_4$ as shown in Refs.~\cite{Hayashi:2024yjc, Hayashi:2024psa}.  

\subsubsection{\texorpdfstring{$SU(N)$}{SU(N)} gauge fields on \texorpdfstring{$\mathbb{R}^3\times S^1$}{R3xS1} and the KvBLLY monopoles}

Let us give a quick review for the $SU(N)$ gauge fields on $\mathbb{R}^3\times S^1$ to be self-contained. 
The $4$d YM theory on $\mathbb{R}^3\times S^1$ becomes the $3$d gauge field coupled with the Polyakov loop $P_4=\calP\exp(\im \int_{S^1}a_4\diff x_4)$. 
We take the Polyakov gauge that diagonalizes $P_4=\mathrm{dig}(\rme^{\im \phi_1},\ldots, \rme^{\im \phi_N})$. 
At the generic points of the classical moduli, where $\phi_i\not=\phi_j$ for $i\not=j$, the remaining gauge group is reduced to $SU(N)\to U(1)^{N-1}\rtimes S_N$.\footnote{
One may wonder if the use of the Abelianized description on $\mathbb{R}^3\times S^1$ makes any sense for pure YM theory as the one-loop potential prefers the trivial holonomy, where the Abelianized description breaks down. 
Let us emphasize that we employ the Abelianized description to solve the ``classical'' YM self-dual equation on $\mathbb{R}^2\times T^2$ with the asymmetric torus $L_3\gtrsim NL_4$ with the 't~Hooft twist, so we do not need to worry about the one-loop fluctuations at this moment. 
} 
We can perform the gauge fixing of $S_N$ by requiring that $\vec{\phi}=(\phi_1,\ldots, \phi_N)$ lies inside the fundamental Weyl chamber, 
\begin{equation}
    \vec{\alpha}_n\cdot \vec{\phi}>0\,\, (n=1,\ldots,N-1),\quad -\vec{\alpha}_N\cdot \vec{\phi}<2\pi. 
    \label{eq:WeylChamber}
\end{equation}
For our convention of the $SU(N)$ root and weight vectors, see Appendix~\ref{sec:RootWeight}. 
Then, off-diagonal gluons are gapped at generic holonomy backgrounds, and the remaining massless gluons can be dualized to the $U(1)^{N-1}$ compact boson $\vec{\sigma}\sim \vec{\sigma}+2\pi \vec{\mu}_i$ using the $3$d Abelian duality: The $U(1)^{N-1}$ field strength of massless diagonal gluons is given by 
\begin{equation}
    \vec{f}=\frac{\im g^2}{4\pi L_4}\star\left(\diff \vec{\sigma}+\frac{\theta}{2\pi}\diff \vec{\phi}\right). 
    \label{eq:AbelianDuality}
\end{equation}
The effective classical Lagrangian becomes
\begin{align}
    S&=\int\left\{\frac{1}{g^2 L_4}|\diff \vec{\phi}|^2+\frac{g^2}{16\pi^2 L_4}\left|\diff\left(\vec{\sigma}+\frac{\theta}{2\pi}\vec{\phi}\right)\right|^2\right\} \notag\\
    &=\int\frac{g^2}{16\pi^2 L_4}|\diff \vec{z}|^2, 
\end{align}
where we define
\begin{equation}
    \vec{z}(\bm{x},x_3)=\im\left\{\vec{\sigma}(\bm{x},x_3)+\left(\frac{\theta}{2\pi}+\frac{4\pi \im}{g^2}\right)(\vec{\phi}(\bm{x},x_3)-\vec{\phi}_c)\right\}. 
\end{equation}
For later convenience, we subtract the center-symmetric holonomy, 
\begin{equation}
    \vec{\phi}_c=\frac{2\pi}{N}\vec{\rho}, 
\end{equation}
with the Weyl vector $\vec{\rho}$. 
When $\vec{\phi}=\vec{\phi}_c$, the Polyakov loop becomes $P_4=C^{-1}$, and the choice of the gauge matches with the one~\eqref{eq:ClassicalVacuum} for the $T^2$ compactified setup.

In this Abelianized setup on $\mathbb{R}^3\times S^1$, the $4$d instanton splits into $N$ monopole-instanton constituents~\cite{Lee:1997vp, Lee:1998bb, Lee:1998vu, Kraan:1998kp, Kraan:1998pm, Kraan:1998sn}, which we call KvBLLY monopoles. 
Their key features can be summarized as follows:
\begin{itemize}
    \item They solve the self-dual YM equation on $\mathbb{R}^3\times S^1$. 
    \item They behave as monopoles with the magnetic charge $\vec{\alpha}_n$ for $n=1,\ldots, N$. 
    \item The topological charge of each monopole is given by $Q_{\mathrm{top}}=\frac{1}{N}+\frac{1}{2\pi}\vec{\alpha}_n\cdot (\vec{\phi}_{\infty}-\vec{\phi}_c)$, where $\vec{\phi}_\infty$ is the asymptotic value of $\vec{\phi}(\bm{x},x_3)$ as $(\bm{x},x_3)\to \infty$. 
\end{itemize}
It would deserve to emphasize that there are $N$ fundamental monopoles: $N-1$ of them correspond to the standard 't~Hooft-Polyakov monopoles~\cite{tHooft:1974kcl, Polyakov:1974ek} with the simple roots $\vec{\alpha}_n$ ($n=1,\ldots, N-1$), and there is an extra fundamental monopole whose magnetic charge is given by the Affine simple root $\vec{\alpha}_N$. 
The above information tells these self-dual fundamental monopoles are effectively described by the vertex operators,  
\begin{equation}
    M_n(\bm{x},x_3)=\exp\left(-\frac{8\pi^2}{g^2 N}+\frac{\im \theta}{N}+\vec{\alpha}_n \cdot \vec{z}(\bm{x},x_3)\right). 
    \label{eq:MonopoleVertex}
\end{equation}
The anti-monopole vertices are given by the complex conjugate, 
\begin{equation}
    M^*_n(\bm{x},x_3)=\exp\left(-\frac{8\pi^2}{g^2 N}-\frac{\im \theta}{N}+\vec{\alpha}_n \cdot \vec{z}^{\,*}(\bm{x},x_3)\right).
\end{equation}
As a side remark, we note that these KvBLLY monopoles (and associated bions) are the basic building block for the semiclassical description of confining gauge theories on the $\mathbb{R}^3\times S^1$ setups~\cite{Davies:1999uw, Davies:2000nw, Unsal:2007vu, Unsal:2007jx, Unsal:2008ch, Shifman:2008ja, Poppitz:2008hr, Poppitz:2012sw}.

\subsubsection{Monopole-vortex continuity and the self-dual center vortex}

We now compactify the third direction, 
\begin{equation}
    \mathbb{R}^3\times (S^1)_{L_4} \Rightarrow \mathbb{R}^2\times \underbrace{(S^1)_{L_3}\times (S^1)_{L_4}}_{p/N \text{ flux}}. 
\end{equation}
As long as $L_3\gg N L_4$, we can use the Abelianized description by translating the $p/N$ 't~Hooft flux into the center-twisted boundary condition along $(S^1)_{L_3}$. 

The center transformation $P_4\mapsto \omega P_4$ is naively given by $\vec{\phi}\mapsto \vec{\phi}-2\pi \vec{\mu}_1$, but it brings $\vec{\phi}$ outside of the fundamental Weyl chamber~\eqref{eq:WeylChamber}. 
We need to combine it with the cyclic Weyl permutation $P_W^{-1}$ to maintain the fundamental chamber, and we obtain~\cite{Anber:2015wha} 
\begin{equation}
    \vec{\sigma}\mapsto P_W^{-1} \vec{\sigma},\quad 
    \vec{\phi}\mapsto P_W^{-1} (\vec{\phi}- 2\pi \vec{\mu}_1). 
\end{equation}
This is equivalent with $\vec{z}\mapsto P_W^{-1}\vec{z}$. 
Therefore, introducing the $p/N$ 't~Hooft flux is achieved by taking the center-twisted boundary condition as~\cite{Hayashi:2024yjc, Hayashi:2024psa} 
\begin{equation}
    \vec{z}(\bm{x},x_3)= P_W^{-p}\vec{z}(\bm{x},x_3+L_3). 
    \label{eq:TwistedBC}
\end{equation}

\begin{figure}[t]
\centering
\includegraphics[width = 0.75 \linewidth]{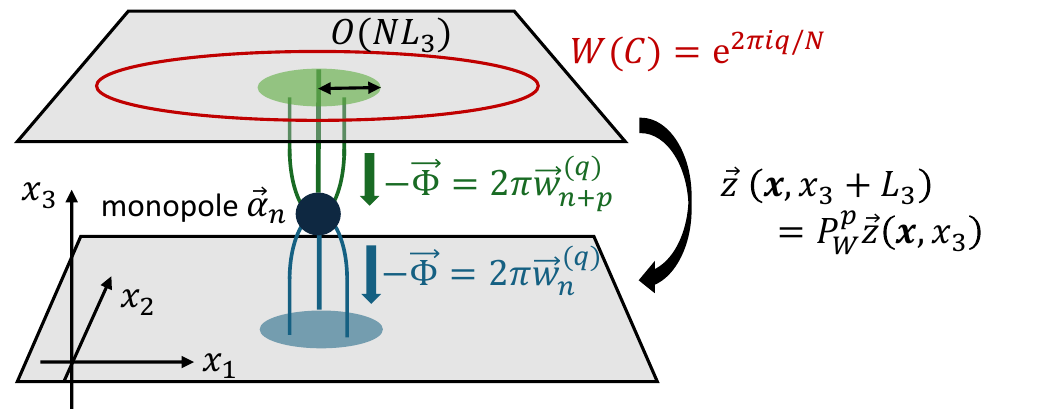}
\caption{
Schematic illustration for the center vortex constructed by a 3d KvBLLY monopole-instanton in $\mathbb{R}^2 \times S^1$ with the center-twisted boundary condition (\ref{eq:TwistedBC}).
The KvBLLY monopole, which carries the magnetic charge $\vec{\alpha}_n$, emits the magnetic flux $\vec{w}_{n}^{(q)}$ and absorbs the magnetic flux $P_W^p\vec{w}_{n}^{(q)}=\vec{w}_{n+p}^{(q)}$.
The magnetic flux is localized on the $x_1$-$x_2$ plane to a size of order $O(NL_3)$.
This localized magnetic flux behaves as a center vortex with flux $q/N$: the Wilson loop acquires the phase $\rme^{\frac{2\pi \im}{N} q}$ when the vortex lies inside the loop. The figure is adapted from the ones in Refs.~\cite{Hayashi:2024yjc, Hayashi:2024psa}.
}
\label{fig:monopole-vortex-schematic}
\end{figure}

With the monopole vertex, $M_n\sim \exp(\vec{\alpha}_n\cdot \vec{z}(\bm{0},x_{3,*}))$, the classical equation of motion becomes 
\begin{align}
    \nabla^2 \vec{\phi} (\bm{x}, x_3) &= 2\pi L_4 \vec{\alpha}_n \delta^{(2)}(\bm{x})\delta(x_3-x_{3,*}), \notag\\
    \nabla^2 \vec{\sigma} (\bm{x}, x_3) &= -2\pi L_4 \vec{\alpha}_n \left(\frac{\theta}{2\pi}+\frac{4\pi\im}{g^2}\right)\delta^{(2)}(\bm{x})\delta(x_3-x_{3,*}). 
\end{align}
We need to solve this with the twisted boundary condition \eqref{eq:TwistedBC} and with $\vec{\sigma}\to 0$ and $\vec{\phi}-\vec{\phi}_c\to 0$ as $|\bm{x}|\to \infty$. 
To solve this, it is useful to note that the (Affine) simple roots can be decomposed as
\begin{equation}
    \vec{\alpha}_n=\vec{w}_{n}^{(q)}-P_W^p \vec{w}_{n}^{(q)}, 
\end{equation}
with 
\begin{equation}
    \vec{w}_{n}^{(q)}:=\vec{\nu}_n+\vec{\nu}_{n+p}+\cdots \vec{\nu}_{n+(q-1)p}.  
\end{equation}
We note that $\vec{w}_n^{(q)}$ belongs to the $q$-box totally antisymmetric representation. 
Then, the solution of the classical equation of motion with the given boundary condition is 
\begin{align}
    &\vec{\phi}(\bm{x},x_3)-\vec{\phi}_c=\frac{\im g^2}{4\pi}\left(\vec{\sigma}(\bm{x},x_3)+\frac{\theta}{2\pi}(\vec{\phi}(\bm{x},x_3)-\vec{\phi}_c)\right) \notag\\
    \qquad &= -\frac{L_4}{2}\sum_{k\in\mathbb{Z}}\vec{w}_{n+pk}^{(q)}\left(\frac{1}{|(\bm{x},x_3-x_{3,*}-kL_3)|}-\frac{1}{|(\bm{x},x_3-x_{3,*}-(k-1)L_3)|}\right). 
    \label{eq:VortexFromMonopole}
\end{align}
By construction, it satisfies the center-twisted boundary condition~\eqref{eq:TwistedBC}. 
To find the asymptotic behavior at $|\bm{x}|\to \infty$, it is useful to take the Fourier transform of this expression,
\begin{align}
    &-\frac{L_4}{2NL_3}\sum_{\ell=0}^{N-1}\vec{w}_{n+p\ell}^{(q)}
    \sum_{m=1}^{\infty}K_0\left(\frac{2\pi}{NL_3}m |\bm{x}|\right) \notag\\
    &\quad \times \left\{\cos\left(\frac{2\pi m}{NL_3}(x_3-x_{3,*}-\ell L_3)\right)-\cos\left(\frac{2\pi m}{NL_3}(x_3-x_{3,*}-(\ell-1) L_3)\right)\right\}, 
    \label{eq:VortexFromMonopole_Fourier}
\end{align}
which explicitly shows that $\vec{\sigma}, \vec{\phi}-\vec{\phi}_c=O(\rme^{-\frac{2\pi}{NL_3}|\bm{x}|})$ as $|\bm{x}|\to \infty$. 
Thus, the magnetic flux forms a localized vortex-like configuration. Moreover, this shows $\vec{\phi}_\infty=\vec{\phi}_c$ and thus the vortex carries $Q_{\mathrm{top}}=\frac{1}{N}+\frac{1}{2\pi}\vec{\alpha}_n\cdot (\vec{\phi}_\infty-\vec{\phi}_c)=\frac{1}{N}$.  
See Figure \ref{fig:monopole-vortex-schematic} for a schematic illustration.

We can check that the $\vec{\alpha}_n$-monopole emits the magnetic flux $\vec{w}_{n}^{(q)}$ in the downward direction and absorbs the flux $P_W^p\vec{w}_{n}^{(q)}=\vec{w}_{n+p}^{(q)}$ from the upward direction: 
Using the duality map~\eqref{eq:AbelianDuality}, the magnetic flux $\vec{\Phi}$ through the $x_1$-$x_2$ plane is expressed as 
\begin{align}
    \vec{\Phi}(x_3)
    &=\int_{\mathbb{R}^2}\diff^2\bm{x} \frac{\im g^2}{4\pi L_4}\partial_3 \left(\vec{\sigma}(\bm{x},x_3)+\frac{\theta}{2\pi}\vec{\phi}(\bm{x},x_3)\right) \notag\\
    &=\pi\sum_{k}\vec{w}_{n+pk}^{(q)}\sum_{k\in \mathbb{Z}}\left[\mathrm{sign}(x_3-x_{3,*}-k L_3)-\mathrm{sign}(x_3-x_{3,*}-(k-1)L_3)\right] \notag\\
    &=\left\{\begin{array}{cl}
        -2\pi\vec{w}_{n}^{(q)} & \quad (0\le x_3<x_{3,*}), \\
        -2\pi\vec{w}_{n+p}^{(q)} & \quad(x_{3,*} < x_{3} < L_{3}). 
    \end{array}\right.
    \label{eq:VortexMagneticFlux}
\end{align}
Thus, this vortex affects the fundamental Wilson loop as follows: 
\begin{align}
    W(C)&= \frac{1}{N}\sum_{k=1}^{N}\exp\left(\im \vec{\nu}_k\cdot \int_C\vec{a}\right)\notag\\
    &= \left\{\begin{array}{cl}
        \exp\left(\frac{2\pi \im}{N} q\right) & \quad (\text{when the vortex lies inside $C$}), \\
        1 & \quad (\text{when the vortex lies outside $C$}), 
    \end{array}\right.
    \label{eq:MagneticFluxOfVortex}
\end{align}
which is nothing but the defining property of the center vortex. 
This gives the explicit construction of the minimal-action center vortex, which carries the magnetic flux $q/N$ with $pq=1$ mod $N$. 
While our construction of the self-dual center vortex is limited to the case $L_3\gg NL_4$ due to the use of the Abelianized effective theory on $\mathbb{R}^3\times S^1$, we believe it should be possible to extend its validity for generic $L_3, L_4$. 

As emphasized in Refs.~\cite{Hayashi:2024yjc, Hayashi:2024psa}, the monopole singularity serves as the junction of the center vortex, which realizes the idea of Refs.~\cite{DelDebbio:1997ke, Ambjorn:1999ym, deForcrand:2000pg} in the semiclassical framework. 
In particular, monopole junctions in the center-vortex networks play the crucial role to have the nontrivial $\theta$ dependence~\cite{Engelhardt:1999xw, Reinhardt:2001kf, Cornwall:1999xw}, and it is consistent with the fact that the center vortex in our setup carries the nonzero topological charge $\pm \frac{1}{N}$ as it secretly contains monopoles inside it.

\subsection{Center-vortex semiclassics and the vacuum structure}
\label{sec:SemiclassicalComputations}

We can now develop the semiclassical description of the confinement by using the dilute gas approximation of center vortices. 
Before going into calculations, let us discuss the validity of the dilute gas approximation~\cite{Tanizaki:2022ngt}. 

We should note that perturbative spectra are gapped for the center-symmetric ground states with the twisted boundary condition. 
Due to the perturbative gap, any localized field excitations have to exponentially decay to the classical vacuum.  
This fact immediately tells 
\begin{itemize}
    \item the center vortex does not have the size moduli, and 
    \item the vortex-(anti)vortex interaction is short ranged. 
\end{itemize}
When $L_3\gg NL_4$, the absence of the size moduli can be confirmed by the explicit construction from the KvBLLY monopoles in \eqref{eq:VortexFromMonopole} and \eqref{eq:VortexFromMonopole_Fourier}.  
Moreover, the vortex-vortex interaction is classically absent due to the self-dual nature, and it is also straightforward to confirm the vortex-antivortex interaction is short ranged. 
But, here let us emphasize the above features are general consequence of the perturbative mass gap. 
Thus, the center-vortex vertex is well-defined without suffering from the infrared divergence. 

In the following, let us set $L_3=L_4=:L$.
The density of the vortex gas is controlled by the compactification scale as its fugacity is dominantly given by 
\begin{equation}
    \exp\left(-\frac{8\pi^2}{Ng^2}\right)=(NL\Lambda)^{\frac{11}{3}}, 
\end{equation}
where we use the result of the $1$-loop renormalized coupling at the scale $\mu=1/(NL)$. 
Thus, we can argue that the vortex gas is dilute when $NL\Lambda\ll 1$. 
Since the vortex-(anti)vortex interaction is short ranged, the ensembles of center vortices can be treated as free gas in this dilute limit.

\subsubsection{Multi-branch vacua and the \texorpdfstring{$\theta$}{theta} dependence}

Let us evaluate the partition function of the center-vortex effective theory. 
As the center vortex has the topological charge $\frac{1}{N}$ and satisfies the self-dual bound, the Boltzmann weight of the single vortex configuration is given by 
\begin{equation}
    K\exp\left(-\frac{8\pi^2}{Ng^2}+\frac{\im \theta}{N}\right), 
\end{equation}
where $K\sim \frac{1}{(NL)^2}$ comes from the fluctuation determinant.\footnote{
The computation of the fluctuation determinant is one of the remaining issues in the center-vortex semiclassics, and the obstacle comes from the fact that the analytic form of the center-vortex configuration is unknown. Our construction from the KvBLLY monopoles in Section~\ref{sec:CenterVortexFromKvBLLY} may be helpful toward its computation, but let us leave it for future works. 
For our purpose of this paper, we still need to know its $N$ scaling in the large-$N$ limit with fixed $l=NL\Lambda \ll 1$, and we claim $K=O(N^0)$. 
As $K$ gives the overall factor of the $2$d energy density, its $N$-scaling is identified with the one for the $2$d energy density.  The $4$d vacuum energy density scales as $N^2$ as there are $N^2$ massless gluons, and the $2$d energy density is roughly given by multiplying $L^2=\frac{l^2}{N^2\Lambda^2}$ to the $4$d energy density, which cancels the overall $N^2$ and gives $K=O(1)$. This counting is also consistent with the absence of perturbative massless degrees of freedom. }
To evaluate the partition function on $\mathbb{R}^2\times T^2$, we need to specify the boundary condition at infinities. Here, we also compactify the $\mathbb{R}^2$ direction to a certain closed $2$-manifold $M_2$, so the $4$d spacetime becomes $M_4=M_2\times (T^2)_{p/N \text{ flux}}$. 
Let $V$ be the volume of $M_2$, and we will eventually take the limit $V\to \infty$ after evaluating the partition function and observables, while we keep the size of $T^2$ to be small.

When compactifying $\mathbb{R}^2$ to $M_2$, we assume to take the periodic boundary condition so that the only nontrivial 't~Hooft flux is $n_{34}=p$. 
This requires that the total topological charge has to be quantized as $Q_{\mathrm{top}}\in \mathbb{Z}$. 
The center-vortex gas configuration with $n$ vortices and $\bar{n}$ anti-vortices has the topological charge
\begin{equation}
    Q_{\mathrm{top}}=\frac{n-\bar{n}}{N}, 
\end{equation}
and thus we need the restriction $n-\bar{n}\in N\mathbb{Z}$ when summing up the center-vortex ensembles. We represent this restriction by introducing the Kronecker delta factor, 
\begin{equation}
    \delta_{n-\bar{n}\in N \mathbb{Z}}=\sum_{k=0}^{N-1}\rme^{-\frac{2\pi \im}{N}k(n-\bar{n})}, 
\end{equation}
and the effective partition function $Z_{\mathrm{eff},\theta}$ can be computed as follows~\cite{Tanizaki:2022ngt}: 
\begin{align}
    Z_{\mathrm{eff},\theta}&=\sum_{n,\bar{n}\ge 0}\frac{\delta_{n-\bar{n}\in N\mathbb{Z}}}{n! \bar{n}!} \left(VK\rme^{-\frac{8\pi^2}{Ng^2}+\frac{\im \theta}{N}}\right)^n \left(VK\rme^{-\frac{8\pi^2}{Ng^2}-\frac{\im \theta}{N}}\right)^{\bar{n}} \notag\\
    &=\sum_{n,\bar{n}\ge 0}\sum_{k=0}^{N-1}\frac{\rme^{-\frac{2\pi \im}{N}k(n-\bar{n})}}{n! \bar{n}!}\left(VK\rme^{-\frac{8\pi^2}{Ng^2}+\frac{\im \theta}{N}}\right)^n \left(VK\rme^{-\frac{8\pi^2}{Ng^2}-\frac{\im \theta}{N}}\right)^{\bar{n}}\notag\\
    &=\sum_{k=0}^{N-1}\exp\left[-V\left(-2K \rme^{-\frac{8\pi^2}{Ng^2}}\cos\left(\frac{\theta-2\pi k}{N}\right)\right)\right]. 
    \label{eq:Z_VortexEFT}
\end{align}
Thus, there are $N$-branch structure of the ground states specified by the discrete label $k\in \mathbb{Z}_N$, and each ground state has the energy density 
\begin{align}
    E_k(\theta) &= -2K\rme^{-\frac{8\pi^2}{Ng^2}}\cos\left(\frac{\theta-2\pi k}{N}\right) \notag\\
    &\sim -\Lambda^2(NL\Lambda)^{\frac{5}{3}}\cos\left(\frac{\theta-2\pi k}{N}\right). 
\end{align}
The partition function $Z_{\mathrm{eff},\theta}$ itself has the $2\pi$ periodicity of the $\theta$ angle, but each ground state $E_k(\theta)$ is only $2\pi N$ periodic. 
The first order transition happens when $\theta$ goes across odd multiples of $\pi$. 

\subsubsection{Confining string tensions}

For the above dilute gas computation of the partition function, there is no dependence on the twist $p$ in \eqref{eq:Z_VortexEFT}, and the center-vortex effective theory for non-minimal twist looks completely identical with the case of the minimal twist in Ref.~\cite{Tanizaki:2022ngt}. 
The twist dependence can be observed by computing the string tension of the Wilson loops.

We consider the Wilson loop $W_\calR(C)$ of the irreducible representation $\calR$ in $M_2 (\to \mathbb{R}^2)$, and then the center vortex acquires an extra phase by \eqref{eq:MagneticFluxOfVortex} if it lies inside $W_\calR(C)$: 
\begin{align}
    \text{Boltzmann weight}=
    \left\{\begin{array}{cl}
       K\exp\left(-\frac{8\pi^2}{Ng^2}+\frac{\im (\theta+2\pi q|\calR|)}{N}\right) &  \, (\text{if the vortex is inside $C$}),\\
       K\exp\left(-\frac{8\pi^2}{Ng^2}+\frac{\im \theta}{N}\right) &  \, (\text{if the vortex is outside $C$}).
    \end{array}\right.
\end{align}
Here, $|\calR|$ is the $N$-ality of $|\calR|$, i.e., the number of boxes of the Young tableaux in mod $N$.  
Let $\calA=\mathrm{Area}(C)$, then the expectation value of the Wilson loop can be calculated as follows:
\begin{align}
    \langle W_\calR(C)\rangle
    &=\frac{1}{Z_{\mathrm{eff},\theta}}\sum_{n_1,n_2,\bar{n}_1,\bar{n}_2}\frac{\calA^{n_1+\bar{n}_1} (V-\calA)^{n_2+\bar{n_2}}}{n_1! \bar{n}_1! n_2! \bar{n}_2!} \left(K\rme^{-\frac{8\pi^2}{Ng^2}}\right)^{n_1+n_2+\bar{n}_1+\bar{n}_2}\notag\\
    &\quad\quad \times \rme^{\frac{\im \theta}{N}(n_1+n_2-\bar{n}_1-\bar{n}_2)}\rme^{\frac{2\pi \im q |\calR|}{N}(n_1-\bar{n}_1)} \delta_{n_1+n_2-\bar{n_1}-\bar{n}_2\in N\mathbb{Z}} \\
    &=\frac{1}{Z_{\mathrm{eff},\theta}}\sum_{k=0}^{N-1}\rme^{-V E_k(\theta)}\exp\Bigl(-\calA (E_{k}(\theta+2\pi q|\calR|)-E_k(\theta))\Bigr).
\end{align}
where  $n_1  $ and $n_2$ denotes the number of vortices inside and outside of Wilson loop, and similarly $\bar{n}_1$   and $\bar{n}_2$ are the corresponding number for anti-vortices. 
Taking the infinite volume limit $V\to \infty$ for $-\pi<\theta<\pi$, the $k=0$ state is selected for the vacuum and we obtain 
\begin{equation}
    \langle W_\calR(C)\rangle = \exp\Bigl(-(E_{0}(\theta+2\pi q|\calR|)-E_0(\theta))\mathrm{Area}(C)\Bigr).  
\end{equation}
The string tensions are given by 
\begin{align}
    T_\calR(\theta)&=E_{0}(\theta+2\pi q|\calR|)-E_0(\theta) \notag\\
    &=2K \rme^{-\frac{8\pi^2}{Ng^2}}\left[\cos\frac{\theta}{N}-\cos\frac{\theta+2\pi q|\calR|}{N}\right],
\end{align}
and, especially at $\theta=0$, they obey 
\begin{equation}
    T_\calR\sim \Lambda^2(NL\Lambda)^{\frac{5}{3}}\sin^2 \left(\frac{\pi q}{N}|R|\right). 
    \label{eq:StringTensions}
\end{equation}
As a common property of the center-vortex mechanism, the string tension depends only on the $N$-ality $|\calR|$ of the representation $\calR$. 
Let us emphasize that the twist dependence appears on the string tensions, since $q$ is an integer that is determined from $p$ by the relation $pq=1$ mod $N$.

\subsection{Semiclassical realization of the generalized anomaly}
\label{sec:Anomaly}

In this section, let us show that the center-vortex semiclassical theory satisfies the anomaly matching condition for the generalized anomaly between the $\mathbb{Z}_N^{[1]}$ symmetry and the $\theta$ periodicity found in Ref.~\cite{Gaiotto:2017yup}. 

To detect the anomaly, we introduce the background $\mathbb{Z}_N$ $2$-form gauge field $B_{4\rmd}$, which is given by the $\mathbb{Z}_N$-valued $2$-cochain, and we couple it to the $SU(N)$ YM theory.\footnote{
We note that the convention for the normalization of $B_{4\rmd}$ is different by the factor of $\frac{2\pi}{N}$ compared with the one in the previous paper~\cite{Tanizaki:2022ngt}. } 
In the lattice regularized description, the gauged action is nothing but \eqref{eq:WilsonActon_ZN2form}, while $B_{4\rmd}$ no longer has to be restricted to the specific form given there. 
Under the presence of this background field, the gauge-invariant topological charge is quantized as~\cite{vanBaal:1982ag}\footnote{
Here, we assume, for simplicity, that the $4$d spacetime has a spin structure and its integral homology is torsion-free, and $B_{4\rmd}$ is lifted to an element of $\mathbb{Z}$-valued cohomology. Without the torsion-free assumption, we need to replace the formula with Pontryagin square, $P_2(B_{4\rmd})=B_{4\rmd}\cup B_{4\rmd}+B_{4\rmd}\cup_1\delta B_{4\rmd}$. } 
\begin{equation}
    Q_{\mathrm{top}}[B_{4\rmd}]\in -\int\frac{1}{2}B_{4\rmd}\cup B_{4\rmd}+\mathbb{Z}. 
\end{equation}
Under the admissibility condition~\cite{Luscher:1981zq}, this formula is now proven to be true with the finite UV lattice regularization~\cite{Abe:2023ncy}. 
As the fractional shift of the topological charge is independent of dynamical fields, the $4$d YM theory has the following generalized anomaly~\cite{Gaiotto:2017yup}, 
\begin{equation}
    Z_{\theta+2\pi}[B_{4\rmd}]=\exp\left(-\frac{2\pi \im}{N}\int \frac{1}{2}B_{4\rmd}\cup B_{4\rmd}\right) Z_{\theta}[B]. 
    \label{eq:GeneralizedAnomaly}
\end{equation}
To match this anomaly, the multi-branch structure of the $\theta$ vacua is required if we assume confinement~\cite{Gaiotto:2017yup, Tanizaki:2017bam, Kikuchi:2017pcp, Tanizaki:2018xto, Cordova:2019jnf, Cordova:2019uob}: 
For a given generic value of $\theta$, the assumption of unbroken $\mathbb{Z}_N^{[1]}$ symmetry tells its partition function in the infinite-volume limit  should respond to $B_{4\rmd}$ as 
\begin{equation}
    Z_{\theta}[B_{4\rmd}]= \exp\left(-\frac{2\pi\im k}{N}\int \frac{1}{2}B_{4\rmd}\cup B_{4\rmd}\right)Z_{\theta}[0]
\end{equation}
for some $k\sim k+N$. 
The background gauge invariance requires the quantization of $k$, and, for the $k$-th confining vacuum, the deconfined dyonic line is generated by $H(C,\Sigma)W^{-k}(C)$ in the language of the Wilson-'t~Hooft classification~\cite{tHooft:1979rtg, Nguyen:2023fun} (see also \cite{Kapustin:2013qsa, Kapustin:2013uxa, Gukov:2013zka}),\footnote{
Here, let us emphasize that the string tensions of dyonic non-genuine lines are well-defined order parameters even though they are surface operators with boundaries. The surfaces of dyonic non-genuine lines are generators of the $1$-form symmetry, and thus the surface dependence is topological, which prohibits the local counterterm that depends on its area. The string tension cannot be removed by the local renormalization respecting the $1$-form symmetry. 
} where $W(C)$ is the Wilson loop and $H(C,\Sigma)$ is the 't~Hooft loop with $\partial\Sigma=C$. 
The anomaly~\eqref{eq:GeneralizedAnomaly} shows that $k$ has to be shifted to $k+1$ as $\theta\to \theta+2\pi$, but this is impossible without level crossing as $k$ takes only discrete values. 
Thus, we need to introduce the multi-branch structure to satisfy the anomaly matching condition with the confinement states.

Under the $T^2$ compactification with the $p/N$ flux, the background gauge field is decomposed as~\cite{Tanizaki:2022ngt}
\begin{equation}
    B_{4\rmd}=B_{2\rmd}+A_{3}\wedge \frac{\diff x_3}{L_3}+A_4\wedge\frac{\diff x_4}{L_4}+p\frac{\diff x_3\wedge \diff x_4}{L_3L_4}, 
\end{equation}
where $B_{2\rmd}$ and $A_{3,4}$ are independent of the compactified coordinates $x_{3,4}$ and they describe the background gauge field of the symmetry~\eqref{eq:1FormSymmetry_compactification} in the $2$d effective theory. 
Substituting it into the above formula~\eqref{eq:GeneralizedAnomaly}, we obtain that 
\begin{equation}
    Z_{\theta+2\pi}^{(p)}[B_{2\rmd},A_3,A_4] = 
    \exp\left(-\frac{2\pi \im }{N}\int ( p B_{2\rmd} - A_3\cup A_4)\right) Z_{\theta}^{(p)}[B_{2\rmd},A_3,A_4]. 
\end{equation}
Thus, the nontrivial 't~Hooft flux plays an important role to maintain the mixed anomaly between the $2$d $1$-form symmetry and the $\theta$ periodicity~\cite{Tanizaki:2017qhf, Yamazaki:2017dra}.

To reproduce this anomaly, we can introduce the background field dependence into the partition function~\eqref{eq:Z_VortexEFT} of the center-vortex semiclassics as follows: 
\begin{align}
    Z_{\mathrm{eff},\theta}[B_{2\rmd},A_3,A_4]
    &=\sum_{k=0}^{N-1} \exp\left(-\frac{2\pi \im k}{N}\int_{M_2} ( p B_{2\rmd} - A_3\cup A_4)\right) \notag\\
    &\qquad \times \exp\left[-V\left(-2K \rme^{-\frac{8\pi^2}{Ng^2}}\cos\left(\frac{\theta-2\pi k}{N}\right)\right)\right].
\end{align}
When we shift $\theta\to \theta+2\pi$, its effect can be canceled by relabeling $k\to k+1$, which produces the anomalous phase. Therefore, the center-vortex effective theory satisfies the anomaly matching condition. 
The discrete label $k$ is originally introduced to represent the integer quantization of the topological charge for the center-vortex gas, but, interestingly, it now acquires a deep physics meaning related to the Wilson-'t~Hooft classification. 

Let us deconstruct how the anomaly is reproduced in the center-vortex effective theory, and we set $\int_{M_2} B_{2\rmd}=m$ for this purpose. 
By reversing the computation in \eqref{eq:Z_VortexEFT}, we find that it introduces the imbalance between the numbers of vortices and anti-vortices as 
\begin{equation}
    n-\bar{n} = - p m \quad \bmod N. 
\end{equation}
One can understand this relation from two different aspects, and, interestingly, it turns out that those two interpretations are consistent in a nontrivial manner. 
The first aspect comes from the topological charge: As the center vortex carries the fractional topological charge $\frac{1}{N}$, the total topological charge is given by  
\begin{equation}
    Q_{\mathrm{top}}=\frac{n-\bar{n}}{N}\in -\frac{pm}{N}+\mathbb{Z}. 
\end{equation}
The fractional shift on the right hand side matches exactly with the one caused by the 't~Hooft twists, $n_{12}=m$ and $n_{34}=p$. 
The second aspect comes from the magnetic flux: Since $\int_{M_2}B_{2\rmd}=m$ introduces the background fractional magnetic flux in $M_2$, its fractional part has to be neutralized by the magnetic fluxes of center vortices; otherwise, the phase of the $2$d Wilson loop cannot be determined uniquely, which leads to inconsistency. As the center vortex carries $\frac{q}{N}$ magnetic flux, the total magnetic flux in $M_2$ is 
\begin{align}
    \frac{q}{N}(n-\bar{n})+\frac{m}{N}
    &=\frac{1}{N}(-qpm + m) \quad \bmod 1 \notag\\
    &=0 \quad \bmod 1, 
\end{align}
where we use $pq=1$ mod $N$. Thus, it is crucial that the center vortex carries the magnetic flux $q/N$ and the topological charge $1/N$ to be consistent with the anomaly matching.

What about the anomaly for $\int_{M_2} A_3\cup A_4$? 
To be specific, let us set $M_2=T^2$ and consider the case, where $\int_{(S^1)_1}A_3=\int_{(S^1)_2}A_4=1$. This background gauge field introduces the twisted boundary condition,
\begin{align}
    &(P_3,P_4)\mapsto (\omega P_3, P_4) \quad \text{as $x_1\mapsto x_1+L_1$}, \notag\\
    &(P_3,P_4)\mapsto (P_3, \omega P_4) \quad \text{as $x_2\mapsto x_2+L_2$}. 
\end{align}
Since we have chosen the gauge so that $(P_3,P_4)=(S^{-p}, C^{-1})$ in developing the semiclassics, these center transformations have to be combined with $SU(N)$ gauge transformations to maintain $(P_3,P_4)=(S^{-p}, C^{-1})$. 
Thus, we redefine the boundary condition as
\begin{align}
    &(P_3,P_4)\mapsto C^{-q}(\omega P_3, P_4)C^{q} \quad \text{as $x_1\mapsto x_1+L_1$}, \notag\\
    &(P_3,P_4)\mapsto S(P_3, \omega P_4)S^{-1} \quad\,\, \text{ as $x_2\mapsto x_2+L_2$}. 
\end{align}
This boundary condition requires the fractional magnetic flux $q/N$ in the $1$-$2$ plane, and there has to be the imbalance between the numbers of vortices and anti-vortices as $n-\bar{n}=1$ mod $N$. 
That is, the nontrivial twists with the $\mathbb{Z}_N\times \mathbb{Z}_N$ $0$-form center symmetry can be absorbed by the one of the $1$-form center symmetry within the center-vortex effective theory, which is known as the symmetry fractionalization~\cite{Barkeshli:2014cna, Delmastro:2022pfo}, and we here observe its microscopic origin explicitly.  

\section{Center stabilization at large \texorpdfstring{$N$}{N} and Fibonacci sequence}
\label{sec:CenterStabilization}

In the previous section, we develop the semiclassical description of the confinement states for the weak-coupling regimes on small $\mathbb{R}^2\times (T^2)_{p/N \text{ flux}}$. 
It is an important question if this weak-coupling confinement caused by the center-vortex gas is smoothly connected to the strongly-coupled confinement phase on $\mathbb{R}^4$ via decompactification of $(T^2)_{p/N \text{ flux}}$. 
We would like to address this question especially in the large-$N$ limit. 
On the opposite limit, $N=2$, the recent numerical Monte Carlo simulation in Refs.~\cite{Soler:2025vwc, Bergner:2025qsm} suggest the smooth crossover between the weakly-coupled center-vortex gas and the strongly-coupled confinement state. 

We shall consider whether the center-symmetric vacuum obtained by the center-vortex semiclassics is stable in the large-$N$ limit. 
Here, we should note the $4$d $\mathbb{Z}_N$ $1$-form symmetry splits into the $\mathbb{Z}_N$ $1$-form and $\mathbb{Z}_N\times \mathbb{Z}_N$ $0$-form symmetries in the $2$d effective theory, as discussed in \eqref{eq:1FormSymmetry_compactification}. 
When talking about the large-$N$ stability of the center symmetry, we must discuss it both from the viewpoint of the $1$-form and $0$-form center symmetries on the $\mathbb{R}^2\times T^2$ setup. 
Let us discuss them separately in the following.

\subsection{Large-\texorpdfstring{$N$}{N} (in)stability of \texorpdfstring{$0$}{0}-form center-symmetric vacuum}
\label{sec:0FormCenterStabilization}

As mentioned in the introduction, the large-$N$ center (in)stability is extensively discussed in the context of the TEK model~\cite{GonzalezArroyo:2010ss, Guralnik:2002ru, Bietenholz:2006cz, Teper:2006sp, Azeyanagi:2007su, Chamizo:2016msz, GarciaPerez:2018fkj, Bribian:2019ybc}, and we note that the question here is about the stability of $\mathbb{Z}_N\times \mathbb{Z}_N$ $0$-form center symmetry in our context. 
We first discuss this problem from the tachyonic instability of the gluon propagator and next discuss the problem from the action-versus-entropy argument. 

Let us start with giving a brief review of Ref.~\cite{Guralnik:2002ru} by Guralnik et al. in our context. 
In Section~\ref{sec:classical}, we find the center-symmetric classical vacuum in \eqref{eq:ClassicalVacuum} and the perturbative spectrum of gluons in \eqref{eq:PerturbativeGap}. 
Due to the perturbative mass gap, we have claimed the particle-like excitations can be integrated out without infrared divergences and obtain the $\mathbb{Z}_N\times \mathbb{Z}_N$ center symmetric state (with broken $(\mathbb{Z}_N^{[1]})_{2\rmd}$ symmetry). 
However, the number of fluctuations grows as $N^2$. It this classical stability robust even in the large-$N$ limit? 

We need to perform the explicit one-loop computation of the gluon self-energy with the classical vacuum~\eqref{eq:ClassicalVacuum} to answer this question, and this is done in Ref.~\cite{Guralnik:2002ru} for the $U(1)$ gauge theory on non-commutative torus, which is Morita equivalent to the current setup except that the $SU(N)$ gauge group is replaced by $U(N)$. 
To understand the result, let us consider the one-loop graph for the self-energy coming from the four-point vertex $\tr([a_\mu,a_\nu]^2)$. 
As the $\mathfrak{su}(N)$ basis $J_{\vec{\ell}}$ given in \eqref{eq:SUN_clockshiftbasis} satisfies $J_{\vec{\ell}}J_{\vec{k}}=\omega^{\frac{q}{2}\vec{\ell}\times \vec{k}}J_{\vec{\ell}+\vec{k}}$ with $\vec{\ell}\times \vec{k}=\varepsilon_{IJ}\ell_I k_J$, 
the color trace of the four-point vertex becomes 
\begin{equation}
    \tr(J_{\vec{\ell}_1}J_{\vec{\ell}_2}J_{\vec{\ell}_3}J_{\vec{\ell}_4})=N \delta_{\vec{\ell}_1+\vec{\ell}_2+\vec{\ell}_3+\vec{\ell}_4,\vec{0}}
    \exp\left(\frac{\pi \im q}{N}\sum_{1\le i<j\le 4}\vec{\ell}_i\times \vec{\ell}_j\right). 
    \label{eq:ColorTrace_4pt}
\end{equation}
Two of the four legs have to be closed into an internal loop to obtain the self-energy diagram, and there are two classes of such diagrams: 
\begin{enumerate}
    \item The planar diagram is obtained by contracting two consecutive legs, say $\vec{n}:=\vec{\ell}_2=-\vec{\ell}_3$, and then $\vec{\ell}:=\vec{\ell}_1=-\vec{\ell}_4$ correspond to external color-momentum. The oscillatory factor in \eqref{eq:ColorTrace_4pt} vanishes. 
    \item The non-planar diagram is obtained by contracting non-consecutive legs, say $\vec{n}=\vec{\ell}_2=-\vec{\ell}_4$, and $\vec{\ell}=\vec{\ell}_1=-\vec{\ell}_3$ is the external color-momentum. 
    The oscillatory factor in \eqref{eq:ColorTrace_4pt} survives as $\exp\left(\frac{2\pi\im q}{N}\vec{\ell}\times \vec{n}\right)$. 
\end{enumerate}
In particular, the non-planar contribution gives a negative contribution for the self-energy and lowers the perturbative mass gap. 
If $|q\vec{\ell}|\ll N$ at large $N$, the phase factor does not oscillate almost at all, and the non-planar contribution becomes larger and dominates the others. 
As a result, the effective mass gap of transverse gluons $a_{\mu=3,4}$ with $|q\ell|\ll N$ is changed from the tree-level result~\eqref{eq:PerturbativeGap} as \cite{Guralnik:2002ru} 
\begin{equation}
    m_{\mathrm{eff}}^2=\left(\frac{2\pi}{NL}\right)^2 \left(|\vec{\ell}|^2-\# g^2 \frac{N^2}{|q\vec{\ell}|^2}\right). 
    \label{eq:OneLoopMass}
\end{equation}
Since the transverse modes $a_{\mu=3,4}$ correspond to the holonomy degrees of freedom, this suggests tachyonic instability for the $\mathbb{Z}_N\times \mathbb{Z}_N$-symmetric vacuum at some $Ng^2_*\sim N^{-1}$. 
Ref.~\cite{Bietenholz:2006cz} performs the Monte Carlo simulation with $p=\frac{N-1}{2}$ with odd $N$ instead of $p=1$ and also confirms spontaneous center breaking at intermediate couplings nonperturbatively, but the phase transition turns out to occur at a weaker coupling, $Ng^2_*\sim N^{-2}$. 
This weak-coupling phase transition occurs at a length scale $L_{\rm weak} = 
\rme^{-O(N^2)  } \Lambda^{-1}$ shown in Fig.~\ref{fig:phases}(b).

The one-loop self-energy computation already gives us a hint to achieve the center stabilization: 
If we can make the oscillatory factor $\exp\left(\frac{2\pi\im q}{N}\vec{\ell}\times \vec{n}\right)$ operative even at large $N$, the non-planar contributions are suppressed and do not cause tachyonic instability. 
This requires to use $N$-dependent 't~Hooft twist, $p\rightsquigarrow p(N)$, because it turns out that the phase factor stops oscillating at large-$N$ when $\vec{\ell}$ is a multiple of $p$ as $pq=1$ mod $N$.\footnote{
Another alternative is to introduce  $n_f$ massless or light adjoint fermions with periodic boundary conditions. In this case, \eqref{eq:OneLoopMass} is modified into 
$m_{\mathrm{eff}}^2=\left(\frac{2\pi}{NL}\right)^2 \left(|\vec{\ell}|^2+ (n_f-1) \# g^2 \frac{N^2}{|q\vec{\ell}|^2}\right)$ and the instability completely disappears. 
However, in this case, the theory is changed to ${\cal N}=1$ SYM or QCD(adj), and we do not consider this possibility here. For the center-vortex semiclassics of those cases, see Refs.~\cite{Tanizaki:2022ngt, Hayashi:2024psa}. 
} 

The another but related aspect of the large-$N$ center breaking comes out from the action-versus-entropy argument, as given in Ref.~\cite{Teper:2006sp} for the one-site matrix model:  
While the center-symmetric twist eater is the absolute minimum of the classical action, singular torons also become local minima for $N\gg 1$ and the difference of the classical actions between the twist eater and singular torons is $O(N/(Ng^2))$ for the minimal twist,\footnote{
The similar analysis for the $\mathbb{R}^2\times (1^2 \text{ twisted lattice})$ shows the classical-action difference between the center-symmetric and broken vacua is $O((Ng^2)^{-1})$ [see Appendix~\ref{sec:ActionvsEntropy}]. 
Equating it with the $O(N^2)$ entropy factor estimates the phase transition point as $Ng_*^2\sim N^{-2}$, which reproduces the numerical observation in Ref.~\cite{Bietenholz:2006cz}; the first-order transition occurs before tachyonic instability of the center-symmetric vacuum.
} which is not necessarily strong enough to compete with $O(N^2)$ entropy factor. 
Ref.~\cite{GonzalezArroyo:2010ss} then discusses the condition to obtain the classical-action gain of $O(N^2/(Ng^2))$ and also the condition to make singular torons locally unstable, which results in the proposal to use the $N$-dependent twist for the large-$N$ center stabilization. 
In Appendix~\ref{sec:ActionvsEntropy}, we give the action-versus-entropy discussion for the case of our $\mathbb{R}^2\times T^2$ setup. 

Choosing the optimal pairs $(N,p)$ to suppress non-planar contributions now becomes the number-theoretic problem,\footnote{
Let us emphasize two stability conditions come out from the tachyonic-instability argument and from the action-versus-entropy argument, and they basically lead the same criterion for $(N,p,q)$. 
The negative shift of the effective mass in Ref.~\cite{Guralnik:2002ru} dominantly comes from the dual lattice momentum closest to the fractional momentum, $(\frac{q}{N}\varepsilon_{IJ}\ell_J)_{I=3,4}$, after Poisson summation. 
If any $O(1)$ integer multiples of $q/N$ are sufficiently far from integers, the mass shift becomes tiny and does not cause the tachyonic instability.  

In the action-versus-entropy argument, the stability of the twist eater need the $O(N^2)$ gain of the classical action compared with not only singular torons but also partially center-broken vacua. Corresponding analysis on $\mathbb{R}^2\times T^2$ requires any $O(1)$ multiples of $p/N$ are far from integers [see Appendix~\ref{sec:ActionvsEntropy}]. 

To see the equivalence of two requirements, let us discuss the contraposition; if some $O(1)$ multiple of $p/N$ is close to an integer, then there exists some $O(1)$ multiple of $q/N$ close to an integer. Let us take some $n=O(1)$ such that $n \frac{p}{N}=\frac{p'}{N}$ mod $1$ with $p'=O(1)$ by assumption. 
We then find that $p'\frac{q}{N}=\frac{n}{N}pq=\frac{n}{N}$ mod $1$, and this is what we want as $p',n=O(1)$. 
} and this is investigated in the Hamiltonian formalism for the $2+1$d YM theory in Ref.~\cite{Chamizo:2016msz}, which proposes to use $N=F_{n+2}$ and $p=F_n$, where $F_n$ is the Fibonacci sequence. 
The point is that we need to choose the sequence of $(N,p,q)$ so that $|q|/N$ converges to an irrational number that is ``difficult to be approximated'' by rational numbers as $N\to \infty$.

Although this condition appears for the large-$N$ stabilization of the $\mathbb{Z}_N\times \mathbb{Z}_N$ $0$-form center-symmetric vacuum, we shall rediscover the same criterion from the large-$N$ center stability of the $1$-form symmetry in Section~\ref{sec:1FormCenterStabilization}.

\subsection{Confining strings at large \texorpdfstring{$N$}{N} and \texorpdfstring{$1$}{1}-form center stabilization}
\label{sec:1FormCenterStabilization}

Let us discuss the large-$N$ behaviors of the confining strings in the semiclassical center-vortex theory to understand the large-$N$ center stabilization in terms of the $2$d $1$-form symmetry on the $\mathbb{R}^2\times T^2$ setup. 
Our following analysis is based on the formula~\eqref{eq:StringTensions}, and let us recapitulate it here for convenience: 
\begin{equation}
    T_\calR = \sigma\sin^2 \left(\frac{\pi q}{N}|R|\right),
\end{equation}
with $\sigma \sim \Lambda^2 (NL\Lambda)^{\frac{5}{3}}$. 

\subsubsection{Large-\texorpdfstring{$N$}{N} deconfinement of Wilson loops for \texorpdfstring{$p=O(1)$}{p=O(1)}}
\label{sec:1FormCenterStabilization-stringtension}
We first consider the case when $p$ is fixed and independent of $N$. For simplicity, we begin with the simplest case, $p=1$ and then $q=1$. 
Then, when the $N$-ality $|\calR|$ of the Wilson loop is $O(1)$, the string tension behaves as 
\begin{equation}
    T_{\calR}\approx \sigma \left(\frac{\pi |\calR|}{N}\right)^2 \to 0 \quad (N\to \infty). 
\end{equation}
Thus, the string tensions vanish in the large-$N$ limit. 
This behavior is quite suggestive: For $p=1$, the confinement force by the center vortex is too weak to ensure its stability.

What about the other choice of $p$ with $p=O(1)$? In the above discussion, the weakness of the confinement force seems to come from the fact that $\frac{q}{N}\to 0$ as $N\to \infty$ for $p=q=1$, and this can be circumvented by other choices of $p$. 
For example, let us consider the case of $N=2M-1$, and then we may take $p=2$, which satisfies $\gcd(N,p)=1$. 
Then, $q=M$ satisfies $pq=1$ mod $N$, and $\frac{q}{N}\to \frac{1}{2}$ as $N\to \infty$. 
In this case, the fundamental Wilson loop (or more generally Wilson loop with odd charges) has the non-vanishing string tension, 
\begin{equation}
    T_{\Box} \approx \sigma. 
\end{equation}
This looks good at the first thought, but it is insufficient to ensure the stability. 
Let us consider the $2$-box representation, then its string tension is 
\begin{align}
    T_{|\calR|=2}
    &= \sigma \sin^2\left(\frac{\pi M\cdot 2}{2M-1}\right) \notag\\
    &\approx \sigma \left(\frac{\pi}{2M-1}\right)^2 \to 0 \quad (M\to \infty). 
\end{align}
Therefore, the large-$N$ deconfinement takes place for the even-box representations, and the $2$d $1$-form symmetry is spontaneously broken at $N\to\infty$.

Since $pq=1$ mod $N$, the string tension for $|\calR|=p$ always vanishes in the large-$N$ limit: 
\begin{align}
    T_{|\calR|=p}
    &= \sigma \sin^2 \left(\frac{\pi qp}{N}\right) \notag\\
    &\approx \sigma \left(\frac{\pi}{N}\right)^2 \to 0 \quad (N\to \infty). 
\end{align}
Thus, for any choice of $(N,p)$, there exist confining strings with the tension of $O(N^{-2})$. 
For the large-$N$ center stabilization, we need to reshuffle the spectrum in a clever way and assign those ``almost deconfining'' strings to heavy observables in the large-$N$ limit.

The above observation suggests the criterion for the large-$N$ center stabilization. 
We require that the string tension \eqref{eq:StringTensions} for ``light'' representation $|\calR| \sim O(1)$ does not vanish, that is,
\begin{itemize}
    \item All $O(1)$ multiples of $q$ are $O(N)$ in mod $N$. 
\end{itemize}
From $pq=1$ mod $N$, this condition implies that any $O(1)$ multiples of $p$ should not become $O(1)$ mod $N$ either.
The large-$N$ deconfined strings appear for $|\calR|={pk}$ with $k=O(1)$. This criterion requires that all of them are $O(N)$ observables so that it becomes practically difficult to create those deconfined excitations. 

Let us point out that this condition is identical with the one obtained from the action-versus-entropy argument~\cite{GonzalezArroyo:2010ss, Chamizo:2016msz} after suitable translation between the setups. See also Appendix~\ref{sec:ActionvsEntropy}. 
Thus, the condition for the $1$-form center stability turns out to be identical with the one for the $0$-form center stability. 

\subsubsection{Large-\texorpdfstring{$N$}{N} confinement for \texorpdfstring{$N=F_{n+2}$ and $p=F_n$}{N=Fn+2 and p=Fn}}

The studies of $2+1$d YM theory on $\mathbb{R}\times T^2$ with 't~Hooft twist in Refs.~\cite{Chamizo:2016msz, GarciaPerez:2018fkj, Bribian:2019ybc} uncover the particularly good choice for $N$ and $p$, which is given by using the Fibonacci sequence $F_n$ as 
\begin{equation}
    N=F_{n+2}, \quad p=F_n.  
    \label{eq:FibonacciProposal}
\end{equation}
Let us remind ourselves Fibonacci sequence very briefly. Fibonacci sequence is defined by 
\begin{equation}
    F_0=0,\quad F_1=1,\quad 
    F_{n+2}=F_{n}+F_{n+1},  
    \label{eq:FibonaccieSequence}
\end{equation}
and it is convenient to rewrite this expression in the matrix form, 
\begin{equation}
    \begin{pmatrix}
        F_{n+2} & F_{n+1}\\
        F_{n+1} & F_{n}
    \end{pmatrix} 
    = 
    \begin{pmatrix}
        1 & 1\\
        1 & 0
    \end{pmatrix}
    \begin{pmatrix}
        F_{n+1} & F_{n}\\
        F_{n} & F_{n-1}
    \end{pmatrix}
    =
    \begin{pmatrix}
        1 & 1\\
        1 & 0
    \end{pmatrix}^{n+1}. 
\end{equation}
Taking the determinant of the both sides gives $F_{n+2}F_n-F_{n+1}^2=(-1)^{n+1}$. Substituting $F_{n+1}=F_{n+2}-F_n$, we find that 
\begin{equation}
    q=(-1)^n F_n
\end{equation}
satisfies $pq=1$ mod $N$. This also shows $\gcd(N,p)=1$ for $N=F_{n+2}$ and $p=F_n$.\footnote{Using \eqref{eq:FibonaccieSequence}, the Euclidean algorithm shows $\gcd(F_{n+2},F_{n})=\gcd(F_{n+1},F_n)=\cdots=\gcd(F_{2},F_1)=1$.  } 
The general term of Fibonacci sequence is given by 
\begin{equation}
    F_{n}=\frac{\varphi^n-(-\varphi)^{-n}}{\sqrt{5}},  
\end{equation}
where $\varphi=\frac{1+\sqrt{5}}{2}\approx 1.618\ldots$ is the golden ratio. 

When we use \eqref{eq:FibonacciProposal}, the string tensions in the limit $n\to \infty$ become
\begin{equation}
    T_{\calR}= \sigma \sin^2\left(\frac{\pi F_n |\calR|}{F_{n+2}}\right) \to  \sigma \sin^2\left(\frac{\pi |\calR|}{1+\varphi}\right). 
\end{equation}
As $\frac{1}{1+\varphi}=2-\varphi$, we can rewrite the large-$N$ formula as 
\begin{equation}
    T_\calR = \sigma \sin^2 \left(\pi \varphi |\calR|\right). 
\end{equation}
This never vanishes for any $|\calR|$ since $\varphi$ is an irrational number, and thus any $O(1)$ representations feel the linear confinement in the large-$N$ limit. 
Thus, the $2$d $1$-form symmetry is preserved at large-$N$. 

In the above computation, non-vanishing property of large-$N$ string tensions is the consequence of the fact that $|q|/N$ converges to an irrational number. 
What would happen if we consider other sequences satisfying this property instead of Fibonacci sequence? 

For any irrational number $\xi$, we can construct the sequence of $(N,p,q)$ with $pq=1$ mod $N$ so that $|q|/N$ converges to the irrational number in a systematic way: We firstly express the irrational number by the simple continued fraction, $\xi=[a_0;a_1,a_2,\ldots]$, and secondly take its finite truncation. 
For the continued fraction, see Appendix~\ref{sec:ContinuedFraction}. 
By truncating the continued fraction as $[a_0;a_1,\ldots, a_n]=:\frac{c_n}{d_n}$, we obtain the pair of integers that satisfies $\gcd(c_n,d_n)=1$, and we may assign $q=\pm c_n$ and $N=d_n$ to find a sequence satisfying the above property; $p$ is given by $p=\pm (-1)^{n-1}d_{n-1}$. 
The proposal of using Fibonacci sequence is obtained by considering $\frac{1}{1+\varphi}=[0;2,1,1,\ldots]$, and we find $c_n=F_n$ and $d_n=F_{n+2}$ for this case: $N=F_{n+2}$, $q=(-1)^nF_{n}$, and $p=-d_{n-1}=-F_{n+1}\equiv F_n$ mod $N$.  

One may wonder if there is any benefit of using the Fibonacci sequence compared with others. 
Let us assume that we consider the limit $|q|/N \to \xi$ with an irrational number $\xi$.  
Finite truncations of the simple continued fraction gives Diophantine approximations of the irrational number, and its quality of $k$-th approximation is controlled as 
\begin{equation}
    \left|\xi-\frac{c_k}{d_k}\right| < \frac{1}{a_{k+1}}\frac{1}{d_k^2}. 
\end{equation}
Then, the string tension for $|\calR|=d_k$ behaves as 
\begin{align}
    T_{|\calR|=d_k} 
    &= \sigma \sin^2 \left(\pi \xi d_k\right) \notag\\
    &\lesssim \sigma \sin^2 \left(\pi \left(c_k \pm \frac{1}{a_{k+1}d_k}\right)\right) \notag\\
    &\approx \sigma \left(\frac{\pi}{a_{k+1}d_k}\right)^2. 
\end{align}
The suppression by $1/d_k^2$ cannot be evaded since the continued fraction expansion provides an infinite series of quadratically good approximations for any irrational numbers. 
However, we should note that there is also a suppression by $1/a_{k+1}^2$, and thus string tensions become parametrically smaller if $a_{k+1}\gg 1$, 
which could cause the potential instability of the center symmetry for the charges $|\calR|=d_k$ even if $|d_k|\ll N=d_n$.  
To circumvent the accidental smallness of string tensions, it is good to use the irrational number whose continued fraction only contains small positive integers. 

We can see that string tensions do not become accidentally small in the case of \eqref{eq:FibonacciProposal}. 
The simple continued fraction of the golden ratio is given by $\varphi=[1;1,1,\ldots]$, and Diophantine approximations are obtained by truncating it as 
\begin{equation}
    \varphi\approx [\underbrace{1;1,1,\ldots, 1}_{k}]=\frac{F_{k+1}}{F_k}.
\end{equation}
The error of this approximation is controlled as $\frac{1}{\sqrt{5}F_k^2}$, i.e., $F_k^2 \left|\varphi-\frac{F_{k+1}}{F_k}\right|\to \frac{1}{\sqrt{5}}$ as $k\to \infty$.  
Therefore, if we set $|\calR|=F_k$ for some $k\ge 1$, we obtain 
\begin{equation}
    T_{|\calR|=F_k}\approx \sigma \sin^2 \left(\frac{\pi}{\sqrt{5}F_k}\right).  
\end{equation}
This shows that we do not encounter accidental suppression for the string tensions beyond unavoidable ones due to the presence of Diophantine approximation. 
We note that quadratic irrationals generally have the similar property, so use of their continued fractions may provide other systematic alternatives of $(N,p,q)$ for large-$N$ center stabilization.

\subsection{Hamiltonian view of the \texorpdfstring{large-$N$}{large-N} confinement on \texorpdfstring{$\mathbb{R} \times S^1 \times (T^2)_{p/N \text{ flux}}$}{RxS1x(T2)p/N flux}}

Lastly, we give another viewpoint on the $2$d $1$-form center stability by a further $S^1$ compactification : $\mathbb{R} \times S^1 \times (T^2)_{p/N~\mathrm{flux}}$. 
We regard the $\mathbb{R}$ direction as the time direction and consider the Hamiltonian viewpoint of this system. 
With the additional $S^1$ compactification, we can observe the stability of the $2$d 1-form symmetry from the stability of the center symmetry acting on the Polyakov loop of this $S^1$, which we call $P_2 = P(S^1)$.
This setup was discussed in Section 2.5 of \cite{Tanizaki:2022ngt}, and it is deeply related to the setup of Ref.~\cite{Yamazaki:2017ulc}.

Let $L$ denote the size of $T^2$ and $L_2$ denote the size of $S^1$, and we impose the following hierarchy, 
\begin{equation}
    \Lambda^{-1} \gg L_2 \gtrsim N L, 
\end{equation}
to see the $2$d $1$-form stability from a quantum-mechanical viewpoint within semiclassical analysis. 
The classical vacuum should satisfy $F_{IJ}=0$ for $I,J=2,3,4$, and we need to look for the holonomies consistent with the boundary condition. 
Due to the 't~Hooft flux along the $3$-$4$ direction, we can take $P_3=S^{-p},~P_4=C^{-1}$ up to gauge transformations.
In this vacuum, the flatness conditions $F_{23}=F_{24}=0$ implies $[P_2,P_3] = [P_2,P_4]= 0$, which give $N$ classical vacua $P_2 = \rme^{\frac{2 \pi \im}{N}m} \mathbf{1}_{N \times N}$ $(m=0,\cdots,N-1)$.
In other words, we have the quantum mechanics with $N$ states $\{ \ket{m} \}_{m=0,\cdots,N-1}$, representing the classically center-broken states, 
\begin{equation}
    \frac{1}{N} \operatorname{tr}(P_2) \ket{m} = \rme^{\frac{2 \pi \im}{N}m} \ket{m}.
\end{equation}

At finite $N$, the fractional instantons, or center vortices, restores the center symmetry for $P_2$. 
We would like to note that the size of the center vortex is roughly given by $NL$, where $L$ is the size of $T^2$. 
Therefore, for $L_2 \gtrsim N L$, the fractional instanton in the effective quantum mechanics is nothing but the center vortex in the 2d effective theory itself.
Since the center vortex rotates the phase of the Wilson loop by $\rme^{\frac{2 \pi \im}{N}q}$, the center vortex gives the tunneling event from $\ket{m}$ to $\ket{m+q}$. 
The transition amplitude up to one fractional (anti-)instanton reads,\footnote{
For this expression, we implicitly assume $\tau\gg NL$ so that the size of the center vortex is negligible and also $\tau L_2 K\rme^{-\frac{8\pi^2}{Ng^2}}\ll 1$ to justify the one-fractional-instanton approximation. 
We note that such $\tau$ exists as we fix $NL\Lambda\ll 1$ even when taking the large-$N$ limit. This is called abelian (or non 't Hooftian) large-$N$ limit, and take place in weakly coupled confining regime on $\mathbb R^3 \times S^1$ \cite{Poppitz:2011wy}. A similar phenomena also occurs in the large-$N$ limit of Seiberg-Witten solution on   $\mathbb R^4$ \cite{Douglas:1995nw}.
} 
\begin{align}
    \braket{m'|\rme^{-\tau \hat{H}_{\mathrm{eff}}}|m} \sim \delta_{m',m} + \tau L_2 K \rme^{-\frac{8\pi^2}{Ng^2}} \left( \rme^{\im \theta /N} \delta_{m', m +q} + \rme^{-\im \theta /N} \delta_{m', m -q} \right),
\end{align}
where $\hat{H}_{\mathrm{eff}}$ represents the Hamiltonian for the effective quantum mechanics, and the indices of the Kronecker deltas are understood in mod $N$.
From this amplitude, we can obtain the effective Hamiltonian for the $N$ states $\{ \ket{m} \}_{m=0,\cdots,N-1}$:
\begin{align}
    \braket{m'|\hat{H}_{\mathrm{eff}}|m} =-L_2 K \rme^{-\frac{8\pi^2}{Ng^2}} \left( \rme^{\im \theta /N} \delta_{m', m +q} + \rme^{-\im \theta /N} \delta_{m', m -q} \right)
\end{align}

The presence of off-diagonal matrix elements means that $|m\rangle$'s are not energy eigenstates. 
The effective Hamiltonian can be diagonalized by the following basis:
\begin{align}
    \widetilde{\ket{k}} := \frac{1}{\sqrt{N}}\sum_{m=0}^{N-1} \rme^{\frac{2 \pi \im k m}{N}p} \ket{m},
\end{align}
with the eigenvalues,
\begin{align}
    \hat{H}_{\mathrm{eff}} \widetilde{\ket{k}} = - 2 L_2 K \rme^{-\frac{8\pi^2}{Ng^2}}  \cos \left( \frac{\theta -2 \pi k}{N} \right) \widetilde{\ket{k}}.
\end{align}
Depending on the $\theta$ angle, one of the states $\widetilde{\ket{k}}$ is chosen as the vacuum.
From the definition of the Fourier-transformed basis, we have  $\frac{1}{N} \operatorname{tr}(P_2)\widetilde{\ket{k}} = \widetilde{\ket{k+q}}$, and therefore the state is center symmetric: 
\begin{align}
   \widetilde{\bra{k}} \operatorname{tr}(P_2^r) \widetilde{\ket{k}} =0, ~~\mathrm{for}~r \neq 0~ (\operatorname{mod}N).
\end{align}
This explains the restoration of center symmetry at finite $N$.

\begin{figure}[t]
\centering
\begin{minipage}{.40\textwidth}
\subfloat[Large $N$ with $p=q=1$]{\includegraphics[width= 0.95\textwidth]{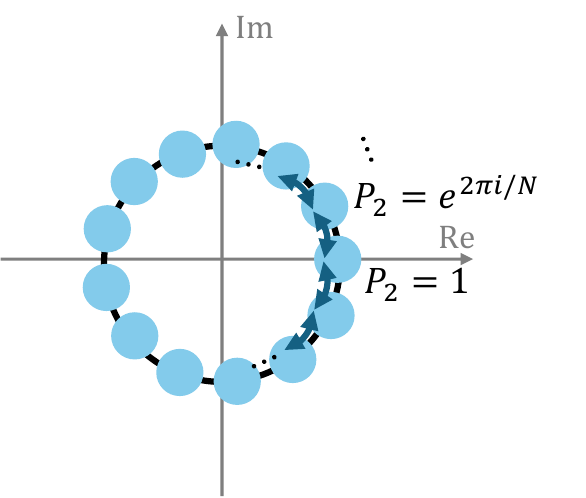}
\label{fig:QMlargeN_minimal}}
\end{minipage}\quad
\begin{minipage}{.40\textwidth}
\subfloat[Large $N$ with $q\sim O(N)$]{\includegraphics[width= 0.95\textwidth]{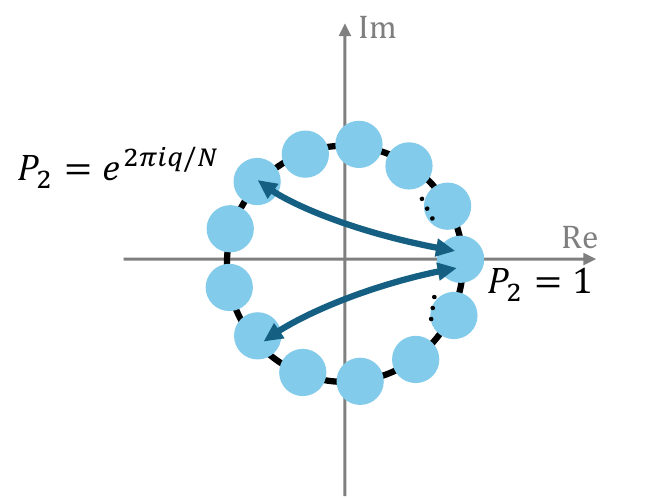}
\label{fig:QMlargeN_nonminimal}}
\end{minipage}
\caption{Schematic picture of how the phase of $P_2$ is randomized by fractional instantons.
(\textbf{a})~At large $N$ with $p=q=1$, fractional instantons rotate the phase of $P_2$ by only $\mathrm{e}^{\pm \frac{2\pi \im}{N}}$.  
Thus, there has to be $O(N)$ events of fractional instantons to randomize the phases of $P_2$, but such processes are suppressed by $\mathrm{e}^{-O(N)}$, signaling that the center symmetry would be spontaneously broken in the large-$N$ limit.
(\textbf{b})~
With $q \sim O(N)$, the phase of the Polyakov loop is randomized by $O(1)$ fractional instantons even in the large-$N$ limit.}
\label{fig:QMfractional}
\end{figure}

Now, we shall discuss whether the ($0$-form) center symmetry for the new $S^1$ is spontaneously broken in the large-$N$ limit or not, which corresponds to the stability of $2$d $1$-form symmetry.
When we refer to the ``large-$N$ symmetry breaking'', the large $N$ limit takes on a role analogous to that of the thermodynamic limit in the standard discussion of spontaneous symmetry breaking.\footnote{ 
The large-$N$ symmetry breaking is detected by the following criterion; 
\begin{align}
    \lim_{\varepsilon \rightarrow +0} \lim_{N \rightarrow \infty} \braket{\operatorname{tr}(P_2^r) }_{\varepsilon} \neq 0, 
    \notag
\end{align}
for some $r\sim O(1)$ with $r \neq 0 ~(\operatorname{mod}N)$, where $\braket{\cdot}_{\varepsilon} $ denotes the vacuum expectation value of the Hamiltonian with symmetry-breaking perturbation,  $\hat{H}_{\mathrm{eff}}^{(\varepsilon)}=\hat{H}_{\mathrm{eff}} - \varepsilon \left( \frac{1}{N}\operatorname{Re} \operatorname{tr}P_2 \right)$. 
}

Figure \ref{fig:QMfractional} shows how fractional instantons try to restore the center symmetry by rotating the phase of the Polyakov loop $P_2$. 
At large $N$ with $p=q=1$, the center symmetry is spontaneously broken in the large-$N$ limit (Figure~\ref{fig:QMlargeN_minimal}).
Here, the fractional instantons provide tunneling events to neighboring classical vacua, but the Polyakov-loop phase is shifted by only $O(1/N)$ through these events.
Thus, to rotate the Polyakov-loop phase by $O(1)$ angles, we need to have $O(N)$ events of fractional instantons, and such processes are $\rme^{-O(N)}$ suppressed.
This situation signals the large-$N$ spontaneous symmetry breaking. 

If we choose the twist so that $q\sim O(N)$, we can avoid this symmetry breaking as the fractional instanton rotates the phase of $P_2$ by $\frac{2\pi q}{N}\sim O(1)$ (Figure~\ref{fig:QMlargeN_nonminimal}). 
The phase of the Polyakov loop can be randomized with only $O(1)$ fractional instantons.
Hence, the Polyakov loop would vanish $\braket{\operatorname{tr}(P_2)} = 0$ even in the large-$N$ limit.

The argument presented here implies the same criterion as in Section \ref{sec:1FormCenterStabilization-stringtension}.
Indeed, if there exists an $O(1)$ multiple $n$ satisfying $nq =O(1)~(\operatorname{mod}N)$, the Polyakov loop to the $n$-th power, $\operatorname{tr} (P_2^n)$, deconfines.
On this quantity, the fractional instanton acts as $\operatorname{tr}(P_2^n) \mapsto \rme^{\frac{2\pi\im}{N}nq} \operatorname{tr}(P_2^n) $, and the multiplying phase factor is only $\rme^{\frac{2\pi\im}{N}nq} = \rme^{O(1/N)}$.
Thus, the situation for $\operatorname{tr} (P_2^n)$ is identical to Figure~\ref{fig:QMlargeN_minimal}.
To rotate its phase by $O(1)$ angles, we need $O(N)$ fractional instantons.
Such amplitudes are $\rme^{-O(N)}$ suppressed, which indicates the large-$N$ center symmetry breaking.
Therefore, for the center stability for $P_2$, we have to choose $q$ such that all $O(1)$ multiples of $q$ are still $O(N)$ in mod $N$.


\acknowledgments
The authors appreciate useful discussion with Georg Bergner, Claudio Bonnano, Aleksey Cherman, Margarita Garc\'\i{}a P\'erez during the conference ``Lattice and Continuum Approaches to Strongly Coupled QFT'' at Kavli Institute for Theoretical Physics (KITP). 
This work was partially supported by Japan Society for the Promotion of Science (JSPS) Research Fellowship for Young Scientists Grant No. 23KJ1161 (Y.H.), by JSPS KAKENHI Grant No. 23K22489 (Y.T.), and by Center for Gravitational Physics and Quantum Information (CGPQI) at Yukawa Institute for Theoretical Physics.  M. \"U. is supported by U.S. Department of Energy, Office 
of Science, Office of Nuclear Physics under Award Number 
  DE-FG02-03ER41260. 

\appendix

\section{Action-vs.-entropy argument for \texorpdfstring{$0$}{0}-form center stability at large \texorpdfstring{$N$}{N} }
\label{sec:ActionvsEntropy}

In this appendix, let us consider the action-versus-entropy argument for our $\mathbb{R}^2\times T^2$ setup to derive the stability criterion for the $0$-form center-symmetric vacuum. 
For simplicity, we approximate $T^2$ by the $1\times 1$ Wilson lattice,\footnote{
Replacement of $T^2$ by a $1\times 1$ lattice only has a  minor effect on the properties of the theory. In particular, it is still true that all modes are perturbatively gapped, but the classical spectrum \eqref{eq:PerturbativeGap} of the continuum theory is replaced with $ m^2_{\ell_3, \ell_4}= \frac{4}{a^2} \left( \sin^2  \frac{\pi  \ell_3}{N}  + \sin^2  \frac{\pi \ell_4}{N}  \right) $. The spectrum is also a realization of color-momentum transmutation in TEK models. Although it is a 1-site model with side-length $a$, the spectrum for $N \gg 1$ becomes the one of a theory on a torus with  effective 
side-length $L_{\rm eff} =Na$.} 
which is equivalent to consider the $2$d YM theory coupled to two $SU(N)$-valued adjoint scalars $P_3,P_4$ with the potential term,
\begin{equation}
    V_{\mathrm{cl}}(P_3,P_4)=\frac{N}{\lambda}
    \tr\left[\left(P_3P_4-\rme^{\frac{2\pi \im p}{N}}P_4 P_3\right)\left(P_3P_4-\rme^{\frac{2\pi \im p}{N}}P_4 P_3\right)^\dagger\right], 
\end{equation}
where $\lambda=Ng^2$ is the 't~Hooft coupling and $p$ corresponds to the 't~Hooft flux for $T^2$. 

To understand which local minimum is selected for the quantum vacuum, we need to evaluate the effective potential including the effect of the fluctuation. Here, let us give a rough criterion to estimate the effective potential for each local minimum as 
\begin{equation}
    V_{\mathrm{eff}}=V_{\mathrm{cl}}- O(\text{the number of massless modes of a given vacuum}). 
\end{equation}
Two typical configurations are the twist eater and the singular torons: 

\begin{enumerate}
    \item \textbf{Center-symmetric vacuum (``twist eater'')}\\
    The twist eater is given by $P_3=S^{-p}$ and $P_4=C^{-1}$ as \eqref{eq:ClassicalVacuum},  
    which gives the absolute minimum of the potential term $V_{\mathrm{cl}}(P_3,P_4)=0$. 
    This is totally center symmetric in terms of the $0$-form center symmetry, and there is no classical zero mode. Thus, we estimate the effective potential for the twist eater as
    \begin{equation}
        V_{\mathrm{eff}}=0.
    \end{equation} 
    
    \item \textbf{Totally center-broken vacuum (``singular toron'')}\\
    Singular torons are given by 
    \begin{equation}
        P_3=\mathrm{diag}(\rme^{\im \psi_1},\ldots, \rme^{\im \psi_N}),\quad 
        P_4=\mathrm{diag}(\rme^{\im \phi_1},\ldots, \rme^{\im \phi_N}),
        \label{eq:SingularToron}
    \end{equation}
    which generically break the $0$-form center symmetry completely. 
    As $P_3P_4P_3^\dagger P_4^\dagger=\bm{1}_{N\times N}$, their classical actions are independent of eigenvalues, 
    \begin{equation}
        V_{\mathrm{cl}}=\frac{4N^2}{\lambda}\sin^2\frac{\pi p}{N}. 
    \end{equation}
    The eigenvalues should be set so that the number of massless modes is maximized. The maximal number $O(N^2)$ is achieved especially when $P_3, P_4\propto \bm{1}_{N\times N}$, and thus  
    \begin{equation}
        V_{\mathrm{eff}}=O\left(\frac{N^2}{\lambda}\sin^2\frac{\pi p}{N}\right) - O(N^2). 
    \end{equation}
\end{enumerate}

Let us consider the case of $p=O(1)$ for $N\gg 1$. Then, the classical action of the singular toron becomes $O(1/\lambda)$. 
In the weak-coupling limit, $\lambda\to 0$, the twist eater is preferred compared with singular torons, but this is reversed around 
\begin{equation}
    \lambda_*= O(N^{-2}). 
\end{equation}
In particular, this suggests the appearance of the first-order transition before having the tachyonic local instability of the center-symmetric vacuum of Ref.~\cite{Guralnik:2002ru}. 

Next, let us consider the case of $p=\frac{N-1}{2}$ with odd $N$, which is the situation of Ref.~\cite{Bietenholz:2006cz}. As $p/N\approx 1/2$, the classical action of singular torons become $O(N^2/\lambda)$, and the transition to completely center-broken vacuum does not occur as long as $\lambda\ll 1$. 
Thus, the phase transition found in Ref.~\cite{Bietenholz:2006cz} is caused by another classical configuration. 
As such a candidate, one may think about other fluxons~\cite{vanBaal:1983eq}, e.g. $P_3=S^{-k}$ and $P_4=C^{-1}$: While their actions become $O(1/\lambda)$ if we choose $p+k\sim O(1)$ in mod $N$, their mass spectra are basically identical with the one for the twist eater up to reshuffling, so their entropy factors are also $O(1)$. 
Therefore, we have to think of other possibilities, and let us propose the following partially center-broken configuration: 

\begin{enumerate}\setcounter{enumi}{2}
    \item \textbf{Partially center-broken state for $p=\frac{N-1}{2}$ with odd $N$}\\
    To construct the partially center-broken configuration, let us decompose $N$ as $N=2M+L$ with both $M,L\sim O(N)$. Our proposal of the candidates is 
    \begin{align}
        P_3 &= [\im \sigma_X\otimes \bm{1}_{M\times M}] \oplus S^{-k}_{(L)}, \notag\\
        P_4 &= \underbrace{[\im \sigma_Z\otimes \bm{1}_{M\times M}]}_{2M\times 2M} \oplus \underbrace{C^{-1}_{(L)}}_{L\times L}, 
        \label{eq:PartiallyBrokenVacuum}
    \end{align}
    where $\sigma_{X,Y,Z}$ is the $SU(2)$ Pauli matrix, and $C_{(L)}$ and $S_{(L)}$ are $SU(L)$ clock and shift matrices. The plaquette is given by
    \begin{equation}
        P_3P_4P_3^\dagger P_4^\dagger = \mathrm{diag}(\underbrace{-1,\ldots, -1}_{2M}, \underbrace{\rme^{\frac{2\pi \im k}{L}}, \ldots, \rme^{\frac{2\pi \im k}{L}}}_{L}). 
    \end{equation}
    Because of the block structure of $P_{3,4}$ and the fact that each sub-block of $P_3P_4P_3^\dagger P_4^\dagger$ is proportional to identity, this configuration solves the classical equation of motion. 
    For $k=(L+1)/2$,  the classical potential can be evaluated as
    \begin{equation}
        V_{\mathrm{cl}}=\frac{2N}{\lambda}\left(4M\sin^2\frac{\pi}{2N}+2L \sin^2\left(\frac{\pi}{2N}-\frac{\pi}{2L}\right)\right). 
    \end{equation}
    As we have assumed $M, L$ are of order $N\gg1$, we find $V_{\mathrm{cl}}=O(\lambda^{-1})$. 
    Since this configuration maintains $SU(M)$ gauge symmetry acting on  $\bm{1}_{M\times M}$ in \eqref{eq:PartiallyBrokenVacuum}, there are $O(M^2)$ massless degrees of freedom, and the effective potential becomes 
    \begin{equation}
        V_{\mathrm{eff}}=O(\lambda^{-1})-O(M^2). 
    \end{equation}
\end{enumerate}

As numerically observed in Ref.~\cite{Bietenholz:2006cz}, we find the first-order phase transition at $\lambda_*\sim O(N^{-2})$ as $M$ is of $O(N)$. 
For the vacuum~\eqref{eq:PartiallyBrokenVacuum}, we find $\tr(P_3)=\tr(P_4)=\tr(P_3P_4)=0$, but non-zero order parameters are, for instance, given by 
\begin{equation}
    \frac{1}{N}\tr(P_3^2)=\frac{1}{N}\tr(P_4^2)=-\frac{2M}{N}\sim O(1), 
\end{equation}
which is also consistent with the result of Ref.~\cite{Bietenholz:2006cz}. The value of $M$ would be able to be estimated by minimizing the $M$ dependence of $V_{\mathrm{eff}}=V_{\mathrm{cl}}-O(M^2)$. 
Although we do not perform it here, let us emphasize the importance of having both $M$, $L=N-2M$ being $O(N)$ at the phase-transition point. 
The proposed configuration itself makes sense also for $M=(N-1)/2$ and $L=1$.
However, when $L=O(1)$, the last $L\times L$ block gives the $O(1)$ contribution for the trace computation, and we end up with $V_{\mathrm{cl}}=O(N/\lambda)$. Such a configuration will not be selected due to this extra factor of $N$ for the classical action, and thus $M$ cannot be too large to ensure that $L=N-2M$ is also large enough in the vicinity of the weak-coupling phase-transition point, $\lambda_*\sim O(N^{-2})$.\footnote{
The relevance of the configuration with $L\sim O(1)$ depends on the magnitude of the coupling $\lambda$. If we are interested in the extremely weak coupling $\lambda \sim O(N^{-2})$ (to examine the center-symmetry-breaking phase transition on this parametrically weak-coupling regions), the configurations with both $M \sim O(N)$, $L=N-2M \sim O(N)$ are relevant.
However, if we are interested in the weak-but-$O(1)$ coupling $\lambda$, the configuration at $M=(N-1)/2$ and $L=1$ would be relevant, because the entropy factor $O(N^2)$ is far larger than the action cost $O(N/\lambda)$ and maximizing the entropy is more important than paying the cost for the action.
Note that the change of a relevant classical configuration (from $L \sim O(N)$ to $L\sim O(1)$) does not necessitate an extra phase transition as no global symmetry breaking is involved.
}

We can easily generalize this consideration for the case of 
\begin{equation}
    \left|\frac{p}{N}-\frac{p'}{n'}\right|=O(N^{-1})
\end{equation}
with some $p',n'=O(1)$. Then, we decompose $N$ as $N=n'M+L$ with both $M,L\sim O(N)$. As a counterpart of \eqref{eq:PartiallyBrokenVacuum}, we consider 
\begin{align}
    P_3 &= [S_{(n')}^{-p'}\otimes \bm{1}_{M\times M}] \oplus S^{-k}_{(L)}, \notag\\
    P_4 &= \underbrace{[C_{n'}^{-1}\,\otimes \bm{1}_{M\times M}]}_{n'M\times n'M} \oplus \underbrace{C^{-1}_{(L)}}_{L\times L}, 
    \label{eq:PartiallyBrokenVacuum_general}
\end{align}
and we can make $V_{\mathrm{cl}}=O(\lambda^{-1})$ for a suitable choice of $k$ so that $\frac{k}{L}\approx \frac{p'}{n'}$ thanks to $L\gg 1$. As there are $M^2$ massless degrees of freedom, we again find the transition for some $\lambda_*\sim N^{-2}$. 
The nontrivial order parameter is given by $\frac{1}{N}\tr(P_3^{n'})=\frac{1}{N}\tr(P_4^{n'})=\pm\frac{n'M}{N}\sim O(1)$. 

To prevent the phase transitions to any patterns of partially center-broken vacua, we must require $p/N$ cannot be approximated by rational numbers with $O(1)$ denominators. 
Equivalently, any $O(1)$ multiples of $p$ should not become $O(1)$ in mod $N$. This is exactly the condition mentioned in the main text for the center-stabilization condition.

\section{Continued fraction and the golden ratio}
\label{sec:ContinuedFraction}

In this appendix, we summarize some properties of the continued fraction, which are useful to understand why the choice $N=F_{n+2}$ and $p=F_n$ is suitable for the center stabilization. 

The (simple) continued fraction $[a_0;a_1,a_2,\ldots]$ expresses a real number $\xi$, 
\begin{equation}
    \xi=a_0+\cfrac{1}{a_1+\cfrac{1}{a_2+\ddots}},  
\end{equation}
where $a_0$ is an integer and $a_1, a_2,\ldots$ are positive integers. In the following, we assume $\xi>0$ and thus $a_0\ge 0$. 
$\xi$ is rational if and only if the continued fraction terminates. 
Thus, we can approximate an irrational number $\xi$ by considering its continued fraction expansion $[a_0;a_1,a_2,\ldots]$ and truncating it as  
\begin{equation}
    \frac{c_n}{d_n}:=[a_0;a_1,a_2,\ldots, a_n]
\end{equation}
with the replacement $a_{n+1}\to\infty$. 
This rational approximation has the following properties:
\begin{itemize}
    \item These rational numbers approximate $\xi$ both from lower and upper sides as 
    \begin{equation}
        \frac{c_0}{d_0}\le \frac{c_2}{d_2}\le \cdots \le \frac{c_{2m}}{d_{2m}} \le \cdots \le 
        \xi 
        \le \cdots \le \frac{c_{2m+1}}{d_{2m+1}}\le \cdots \le \frac{c_3}{d_3} \le \frac{c_1}{d_1}. 
        \label{eq:ContinuedFraction_Prop1}
    \end{equation}
    \item $c_n$ and $d_n$ satisfy the same recursive relation 
    \begin{equation}
        c_n=a_n c_{n-1}+c_{n-2},\quad d_{n}=a_n d_{n-1} + d_{n-2}, 
        \label{eq:ContinuedFraction_Prop2}
    \end{equation}
    with different initial conditions, $c_{-1}=1$, $c_{-2}=0$, and $d_{-1}=0$, $d_{-2}=1$. Thus, 
    \begin{equation}
        \begin{pmatrix}
            c_{n} & c_{n-1}\\
            d_{n} & d_{n-1}
        \end{pmatrix}
        = 
        \begin{pmatrix}
            c_{n-1} & c_{n-2}\\
            d_{n-1} & d_{n-2}
        \end{pmatrix}
        \begin{pmatrix}
            a_{n} & 1 \\
            1 & 0
        \end{pmatrix}
        =
        \begin{pmatrix}
            a_{0} & 1 \\
            1 & 0
        \end{pmatrix}
        \cdots
        \begin{pmatrix}
            a_{n} & 1 \\
            1 & 0
        \end{pmatrix}.
    \end{equation}
    \item These finite truncations satisfy
    \begin{equation}
        \frac{c_{n+1}}{d_{n+1}}-\frac{c_{n}}{d_{n}}=\frac{(-1)^n}{d_n d_{n+1}}. 
        \label{eq:ContinuedFraction_Prop3}
    \end{equation}
\end{itemize}
As a corollary of these properties, the truncations of the continued fraction expansion give quadratically good approximations of the irrational number;
\begin{equation}
    \frac{1}{d_n(d_{n+1}+d_n)}\le \left|\xi-\frac{c_n}{d_n}\right|\le \frac{1}{d_{n}d_{n+1}}. 
    \label{eq:ContinuedFraction_Prop4}
\end{equation}
To obtain this, we note $\left|\frac{c_{n+2}}{d_{n+2}}-\frac{c_n}{d_n}\right|\le \left|\xi-\frac{c_n}{d_n}\right|\le \left|\frac{c_{n+1}}{d_{n+1}}-\frac{c_n}{d_n}\right|$ due to \eqref{eq:ContinuedFraction_Prop1}. 
Using \eqref{eq:ContinuedFraction_Prop3}, we find 
$\left|\frac{1}{d_{n+2}d_{n+1}}-\frac{1}{d_n d_{n+1}}\right|=\frac{d_{n+2}-d_{n}}{d_n d_{n+1}d_{n+2}}\le \left|\xi-\frac{c_n}{d_n}\right|\le \frac{1}{d_{n}d_{n+1}}$. 
For the lower bound, we use the recursive relation \eqref{eq:ContinuedFraction_Prop2} to rewrite it as $\frac{d_{n+2}-d_{n}}{d_n d_{n+1}d_{n+2}}=\frac{a_{n+2}}{d_n(a_{n+2} d_{n+1}+d_n)}\ge \frac{1}{d_n(d_{n+1}+d_n)}$ thanks to $a_{n+2}\ge 1$, and we get the result. 
As $a_{n+1}d_n<d_{n+1}=a_{n+1}d_n+d_{n-1}<(a_{n+1}+1)d_{n}$, the upper and lower bounds may be replaced as 
\begin{equation}
    \frac{1}{a_{n+1}+2}\frac{1}{d_n^2}<\left|\xi-\frac{c_n}{d_n}\right|<\frac{1}{a_{n+1}}\frac{1}{d_n^2},
    \label{eq:ContinuedFraction_Prop5}
\end{equation} 
and the simple continued fraction generates best Diophantine approximations. 
This also shows that a truncation of the continued fraction expansion may give an accidentally good approximation if some of the expansion coefficients $a_{n+1}$ are quite huge.\footnote{As an example, let us consider the continued fraction of $\pi$: $\pi=[3; 7, 15, 1, 292, 1, \ldots]$. Since $a_4=292$ is ``huge'', let us truncate it by replacing $a_4\to \infty$, and we get the famous approximation of $\pi$ as $[3;7,15,1]=\frac{355}{113}$; $|\pi-\frac{355}{113}|\approx 2.67\times 10^{-7}$ is much smaller than $1/113^2$ as it is further suppressed by $1/a_4=1/292$. }

We note that the recursive relation~\eqref{eq:ContinuedFraction_Prop2} has the same form with that of the Fibonacci sequence~\eqref{eq:FibonaccieSequence} when $a_n=1$ for all $n$. That is, the rational approximation of $[1;1,1,\ldots]$ is given by $c_n=F_{n+2}$ and $d_n=F_{n+1}$, and we find 
\begin{equation}
    [1;1,1,\ldots]=\lim_{n\to \infty}\frac{F_{n+2}}{F_{n+1}}=\varphi
\end{equation} 
Moreover, as a consequence of \eqref{eq:ContinuedFraction_Prop4}, we get 
\begin{equation}
    \frac{1}{F_{n+1}F_{n+3}}\le \left|\varphi-\frac{F_{n+2}}{F_{n+1}}\right|\le \frac{1}{F_{n+1}F_{n+2}}. 
\end{equation}
We can improve the upper bound by using the explicit formula of Fibonacci sequence as 
\begin{equation}
    \left|\varphi-\frac{F_{n+2}}{F_{n+1}}\right|< \frac{1}{\sqrt{5} F_{n+1}^2}, 
\end{equation}
giving the bound of Hurwitz's theorem about Diophantine approximations. 
Since all the coefficient in the continued fraction expansion is $a_n=1$ for the golden ratio $\varphi$, its truncation does not become accidentally good, and thus the convergence of the continued fraction expansion is slowest. In this sense, one might be able to say that $\varphi$ is one of the irrational numbers that is most difficult to be approximated by rational numbers. 
We here note that this property is also shared by $\frac{1}{\varphi}=-1+\varphi=[0;1,1,\ldots]$, $\frac{1}{1+\varphi}=2-\varphi=[0;2,1,1,\ldots]$, etc.

\section{Convention of the \texorpdfstring{$SU(N)$}{SU(N)} root and weight vectors}
\label{sec:RootWeight}

We here summarize the convention of the $SU(N)$ root and weight vectors used in this paper. 
Let $\{\vec{e}_n\}_{n=1,\ldots, N}$ be the canonical orthonormal basis of $\mathbb{R}^N$. 
\begin{itemize}
    \item The positive simple roots are given by 
    \begin{equation}
        \vec{\alpha}_n = \vec{e}_n-\vec{e}_{n+1},
    \end{equation}
    with $n=1,\ldots, N-1$. They actually satisfy $\vec{\alpha}_n\cdot \vec{\alpha}_m=2\delta_{nm}-\delta_{n,m\pm 1}$. We also use the Affine root defined by 
    \begin{equation}
        \vec{\alpha}_N=\vec{e}_N-\vec{e}_1=-(\vec{\alpha}_1+\cdots + \vec{\alpha}_{N-1}). 
    \end{equation}
    \item The fundamental weights $\vec{\mu}_n$ ($n=1,\ldots, N-1$) are given by 
    \begin{equation}
        \vec{\mu}_n=\vec{e}_1+\cdots+\vec{e}_n-\frac{n}{N}\sum_{k=1}^{N}\vec{e}_k, 
    \end{equation}
    which satisfy the defining property, $\vec{\mu}_n\cdot \vec{\alpha}_m=\delta_{nm}$ for $n,m=1,\ldots, N-1$. 
    \item The weight vectors in the defining representation are 
    \begin{equation}
        \vec{\nu}_1=\vec{\mu}_1,\, 
        \vec{\nu}_2=\vec{\mu}_1-\vec{\alpha}_1,\, \ldots, \, 
        \vec{\nu}_N=\vec{\mu}_1-\vec{\alpha}_1-\cdots-\vec{\alpha}_{N-1}, 
    \end{equation}
    that is, $\vec{\nu}_n=\vec{e}_n-\frac{1}{N}\sum_k \vec{e}_k$. 
    They satisfy $\vec{\alpha}_n=\vec{\nu}_n-\vec{\nu}_{n+1}$. The following formula is also useful: 
    \begin{equation}
        \vec{\nu}_n\cdot \vec{\nu}_m=-\frac{1}{N}+\delta_{nm}.
    \end{equation} 
    \item We define the cyclic Weyl permutation $P_W$ by $P_W\vec{e}_n=\vec{e}_{n+1}$. 
    It acts on the roots and the weights as 
    \begin{equation}
        P_W \vec{\alpha}_n=\vec{\alpha}_{n+1},\quad P_W \vec{\nu}_n=\vec{\nu}_{n+1}, 
    \end{equation}
    where the subscripts are understood in mod $N$. 
    \item The Weyl vector $\vec{\rho}$ is defined by the half sum of the positive roots, 
    \begin{equation}
        \vec{\rho}=\frac{1}{2}\sum_{i<j}(\vec{e}_i-\vec{e}_j)=\sum_{n=1}^{N-1}\vec{\mu}_n. 
    \end{equation}
    Under the cyclic Weyl permutation, it satisfies $P_W \vec{\rho}=\vec{\rho}-N\vec{\mu}_1$. 
\end{itemize}

\bibliographystyle{utphys}
\bibliography{./QFT.bib, ./refs.bib}

\providecommand{\href}[2]{#2}\begingroup\raggedright\begin{thebibliography}{100}

\bibitem{Wilson:1974sk}
K.~G. Wilson, ``{Confinement of Quarks},''
\href{http://dx.doi.org/10.1103/PhysRevD.10.2445}{{\em Phys. Rev.} {\bfseries D10} (1974) 2445--2459}.

\bibitem{Nambu:1961tp}
Y.~Nambu and G.~Jona-Lasinio, ``{Dynamical Model of Elementary Particles Based on an Analogy with Superconductivity. 1.},''
\href{http://dx.doi.org/10.1103/PhysRev.122.345}{{\em Phys. Rev.} {\bfseries 122} (1961) 345--358}.

\bibitem{Nambu:1961fr}
Y.~Nambu and G.~Jona-Lasinio, ``{Dynamical Model of Elementary Particles Based on an Analogy with Superconductivity. II},''
\href{http://dx.doi.org/10.1103/PhysRev.124.246}{{\em Phys. Rev.} {\bfseries 124} (1961) 246--254}.

\bibitem{tHooft:1979rtg}
G.~'t~Hooft, ``{A Property of Electric and Magnetic Flux in Nonabelian Gauge Theories},''
\href{http://dx.doi.org/10.1016/0550-3213(79)90595-9}{{\em Nucl. Phys.} {\bfseries B153} (1979) 141--160}.

\bibitem{tHooft:1981sps}
G.~{'t Hooft}, ``Aspects of quark confinement,'' \href{http://dx.doi.org/10.1088/0031-8949/24/5/007}{{\em Phys. Scripta} {\bfseries 24} (1981) 841--846}.

\bibitem{Gaiotto:2014kfa}
D.~Gaiotto, A.~Kapustin, N.~Seiberg, and B.~Willett, ``{Generalized Global Symmetries},'' \href{http://dx.doi.org/10.1007/JHEP02(2015)172}{{\em JHEP} {\bfseries 02} (2015) 172},
\href{http://arxiv.org/abs/1412.5148}{{\ttfamily arXiv:1412.5148 [hep-th]}}.

\bibitem{Kapustin:2013uxa}
A.~Kapustin and R.~Thorngren, ``{Higher symmetry and gapped phases of gauge theories},'' \href{http://arxiv.org/abs/1309.4721}{{\ttfamily arXiv:1309.4721 [hep-th]}}.

\bibitem{Kapustin:2014gua}
A.~Kapustin and N.~Seiberg, ``{Coupling a QFT to a TQFT and Duality},'' \href{http://dx.doi.org/10.1007/JHEP04(2014)001}{{\em JHEP} {\bfseries 04} (2014) 001},
\href{http://arxiv.org/abs/1401.0740}{{\ttfamily arXiv:1401.0740 [hep-th]}}.

\bibitem{Gaiotto:2017yup}
D.~Gaiotto, A.~Kapustin, Z.~Komargodski, and N.~Seiberg, ``{Theta, Time Reversal, and Temperature},'' \href{http://dx.doi.org/10.1007/JHEP05(2017)091}{{\em JHEP} {\bfseries 05} (2017) 091},
\href{http://arxiv.org/abs/1703.00501}{{\ttfamily arXiv:1703.00501 [hep-th]}}.

\bibitem{Yamazaki:2017ulc}
M.~Yamazaki and K.~Yonekura, ``{From 4d Yang-Mills to 2d $\mathbb{CP}^{N-1}$ model: IR problem and confinement at weak coupling},'' \href{http://dx.doi.org/10.1007/JHEP07(2017)088}{{\em JHEP} {\bfseries 07} (2017) 088},
\href{http://arxiv.org/abs/1704.05852}{{\ttfamily arXiv:1704.05852 [hep-th]}}.

\bibitem{Cox:2021vsa}
A.~A. Cox, E.~Poppitz, and F.~D. Wandler, ``{The mixed 0-form/1-form anomaly in Hilbert space: pouring the new wine into old bottles},'' \href{http://dx.doi.org/10.1007/JHEP10(2021)069}{{\em JHEP} {\bfseries 10} (2021) 069}, \href{http://arxiv.org/abs/2106.11442}{{\ttfamily arXiv:2106.11442 [hep-th]}}.

\bibitem{Tanizaki:2022ngt}
Y.~Tanizaki and M.~\"Unsal, ``{Center vortex and confinement in Yang-Mills theory and QCD with anomaly-preserving compactifications},'' \href{http://dx.doi.org/10.1093/ptep/ptac042}{{\em PTEP} {\bfseries 2022} (2022) 04A108}, \href{http://arxiv.org/abs/2201.06166}{{\ttfamily arXiv:2201.06166 [hep-th]}}.

\bibitem{Poppitz:2022rxv}
E.~Poppitz and F.~D. Wandler, ``{Gauge theory geography: charting a path between semiclassical islands},'' \href{http://dx.doi.org/10.1007/JHEP02(2023)014}{{\em JHEP} {\bfseries 02} (2023) 014}, \href{http://arxiv.org/abs/2211.10347}{{\ttfamily arXiv:2211.10347 [hep-th]}}.

\bibitem{Shimizu:2017asf}
H.~Shimizu and K.~Yonekura, ``{Anomaly constraints on deconfinement and chiral phase transition},'' \href{http://dx.doi.org/10.1103/PhysRevD.97.105011}{{\em Phys. Rev.} {\bfseries D97} no.~10, (2018) 105011},
\href{http://arxiv.org/abs/1706.06104}{{\ttfamily arXiv:1706.06104 [hep-th]}}.

\bibitem{Tanizaki:2017qhf}
Y.~Tanizaki, T.~Misumi, and N.~Sakai, ``{Circle compactification and 't Hooft anomaly},'' \href{http://dx.doi.org/10.1007/JHEP12(2017)056}{{\em JHEP} {\bfseries 12} (2017) 056},
\href{http://arxiv.org/abs/1710.08923}{{\ttfamily arXiv:1710.08923 [hep-th]}}.

\bibitem{Yamazaki:2017dra}
M.~Yamazaki, ``{Relating 't Hooft Anomalies of 4d Pure Yang-Mills and 2d $\mathbb{CP}^{N-1}$ Model},'' \href{http://dx.doi.org/10.1007/JHEP10(2018)172}{{\em JHEP} {\bfseries 10} (2018) 172},
\href{http://arxiv.org/abs/1711.04360}{{\ttfamily arXiv:1711.04360 [hep-th]}}.

\bibitem{Tanizaki:2017mtm}
Y.~Tanizaki, Y.~Kikuchi, T.~Misumi, and N.~Sakai, ``{Anomaly matching for phase diagram of massless $\mathbb{Z}_N$-QCD},'' \href{http://dx.doi.org/10.1103/PhysRevD.97.054012}{{\em Phys. Rev.} {\bfseries D97} (2018) 054012},
\href{http://arxiv.org/abs/1711.10487}{{\ttfamily arXiv:1711.10487 [hep-th]}}.

\bibitem{Dunne:2018hog}
G.~V. Dunne, Y.~Tanizaki, and M.~\"Unsal, ``{Quantum Distillation of Hilbert Spaces, Semi-classics and Anomaly Matching},'' \href{http://dx.doi.org/10.1007/JHEP08(2018)068}{{\em JHEP} {\bfseries 08} (2018) 068},
\href{http://arxiv.org/abs/1803.02430}{{\ttfamily arXiv:1803.02430 [hep-th]}}.

\bibitem{Yonekura:2019vyz}
K.~Yonekura, ``{Anomaly matching in QCD thermal phase transition},'' \href{http://dx.doi.org/10.1007/JHEP05(2019)062}{{\em JHEP} {\bfseries 05} (2019) 062},
\href{http://arxiv.org/abs/1901.08188}{{\ttfamily arXiv:1901.08188 [hep-th]}}.

\bibitem{Tanizaki:2022plm}
Y.~Tanizaki and M.~\"Unsal, ``{Semiclassics with \textquoteright{}t Hooft flux background for QCD with 2-index quarks},'' \href{http://dx.doi.org/10.1007/JHEP08(2022)038}{{\em JHEP} {\bfseries 08} (2022) 038}, \href{http://arxiv.org/abs/2205.11339}{{\ttfamily arXiv:2205.11339 [hep-th]}}.

\bibitem{Hayashi:2023wwi}
Y.~Hayashi, Y.~Tanizaki, and H.~Watanabe, ``{Semiclassical analysis of the bifundamental QCD on~$\mathbb{R}^2\times T^2$ with \textquoteright{}t Hooft flux},'' \href{http://dx.doi.org/10.1007/JHEP10(2023)146}{{\em JHEP} {\bfseries 10} (2023) 146}, \href{http://arxiv.org/abs/2307.13954}{{\ttfamily arXiv:2307.13954 [hep-th]}}.

\bibitem{Hayashi:2024qkm}
Y.~Hayashi and Y.~Tanizaki, ``{Semiclassics for the QCD vacuum structure through T$^{2}$-compactification with the baryon-\textquoteright{}t Hooft flux},'' \href{http://dx.doi.org/10.1007/JHEP08(2024)001}{{\em JHEP} {\bfseries 08} (2024) 001}, \href{http://arxiv.org/abs/2402.04320}{{\ttfamily arXiv:2402.04320 [hep-th]}}.

\bibitem{Hayashi:2024gxv}
Y.~Hayashi, Y.~Tanizaki, and H.~Watanabe, ``{Non-supersymmetric duality cascade of QCD(BF) via semiclassics on \ensuremath{\mathbb{R}}$^{2}$\texttimes{} T$^{2}$ with the baryon-\textquoteright{}t Hooft flux},'' \href{http://dx.doi.org/10.1007/JHEP07(2024)033}{{\em JHEP} {\bfseries 07} (2024) 033}, \href{http://arxiv.org/abs/2404.16803}{{\ttfamily arXiv:2404.16803 [hep-th]}}.

\bibitem{Hayashi:2024yjc}
Y.~Hayashi and Y.~Tanizaki, ``{Unifying Monopole and Center Vortex as the Semiclassical Confinement Mechanism},'' \href{http://dx.doi.org/10.1103/PhysRevLett.133.171902}{{\em Phys. Rev. Lett.} {\bfseries 133} no.~17, (2024) 171902}, \href{http://arxiv.org/abs/2405.12402}{{\ttfamily arXiv:2405.12402 [hep-th]}}.

\bibitem{Hayashi:2024psa}
Y.~Hayashi, T.~Misumi, and Y.~Tanizaki, ``{Monopole-vortex continuity of ${\mathcal N}=1$ super Yang-Mills theory on $\mathbb{R}^2 \times S^1 \times S^1$ with 't Hooft twist},'' \href{http://arxiv.org/abs/2410.21392}{{\ttfamily arXiv:2410.21392 [hep-th]}}.

\bibitem{Guvendik:2024umd}
C.~G\"uvendik, T.~Schaefer, and M.~\"Unsal, ``{The metamorphosis of semi-classical mechanisms of confinement: from monopoles on \ensuremath{\mathbb{R}}$^{3}$ \texttimes{} S$^{1}$ to center-vortices on \ensuremath{\mathbb{R}}$^{2}$ \texttimes{} T$^{2}$},'' \href{http://dx.doi.org/10.1007/JHEP11(2024)163}{{\em JHEP} {\bfseries 11} (2024) 163}, \href{http://arxiv.org/abs/2405.13696}{{\ttfamily arXiv:2405.13696 [hep-th]}}.

\bibitem{Witten:1980sp}
E.~Witten, ``{Large N Chiral Dynamics},''
\href{http://dx.doi.org/10.1016/0003-4916(80)90325-5}{{\em Annals Phys.} {\bfseries 128} (1980) 363}.

\bibitem{DiVecchia:1980yfw}
P.~Di~Vecchia and G.~Veneziano, ``{Chiral Dynamics in the Large n Limit},'' \href{http://dx.doi.org/10.1016/0550-3213(80)90370-3}{{\em Nucl. Phys. B} {\bfseries 171} (1980) 253--272}.

\bibitem{tHooft:1981bkw}
G.~'t~Hooft, ``{Topology of the Gauge Condition and New Confinement Phases in Nonabelian Gauge Theories},''
\href{http://dx.doi.org/10.1016/0550-3213(81)90442-9}{{\em Nucl. Phys.} {\bfseries B190} (1981) 455--478}.

\bibitem{Witten:1998uka}
E.~Witten, ``{Theta dependence in the large N limit of four-dimensional gauge theories},'' \href{http://dx.doi.org/10.1103/PhysRevLett.81.2862}{{\em Phys. Rev. Lett.} {\bfseries 81} (1998) 2862--2865},
\href{http://arxiv.org/abs/hep-th/9807109}{{\ttfamily arXiv:hep-th/9807109 [hep-th]}}.

\bibitem{Soler:2025vwc}
I.~Soler, G.~Bergner, and A.~Gonzalez-Arroyo, ``{Fractional instantons and Confinement: first results on a $T_2 \times R^2$ roadmap},'' \href{http://dx.doi.org/10.22323/1.466.0408}{{\em PoS} {\bfseries LATTICE2024} (2025) 408}, \href{http://arxiv.org/abs/2502.09463}{{\ttfamily arXiv:2502.09463 [hep-lat]}}.

\bibitem{Bergner:2025qsm}
G.~Bergner, A.~Gonz\'alez-Arroyo, and I.~Soler, ``{A $T_2 \times R^2$ roadmap to Confinement in SU(2) Yang-Mills theory},'' \href{http://arxiv.org/abs/2505.10396}{{\ttfamily arXiv:2505.10396 [hep-lat]}}.

\bibitem{GonzalezArroyo:1982hz}
A.~Gonzalez-Arroyo and M.~Okawa, ``{The Twisted Eguchi-Kawai Model: A Reduced Model for Large N Lattice Gauge Theory},''
\href{http://dx.doi.org/10.1103/PhysRevD.27.2397}{{\em Phys. Rev.} {\bfseries D27} (1983) 2397}.

\bibitem{GonzalezArroyo:1982ub}
A.~Gonzalez-Arroyo and M.~Okawa, ``{A Twisted Model for Large $N$ Lattice Gauge Theory},''
\href{http://dx.doi.org/10.1016/0370-2693(83)90647-0}{{\em Phys. Lett.} {\bfseries 120B} (1983) 174--178}.

\bibitem{GonzalezArroyo:2010ss}
A.~Gonzalez-Arroyo and M.~Okawa, ``{Large $N$ reduction with the Twisted Eguchi-Kawai model},'' \href{http://dx.doi.org/10.1007/JHEP07(2010)043}{{\em JHEP} {\bfseries 07} (2010) 043},
\href{http://arxiv.org/abs/1005.1981}{{\ttfamily arXiv:1005.1981 [hep-th]}}.

\bibitem{Eguchi:1982nm}
T.~Eguchi and H.~Kawai, ``{Reduction of Dynamical Degrees of Freedom in the Large N Gauge Theory},''
\href{http://dx.doi.org/10.1103/PhysRevLett.48.1063}{{\em Phys. Rev. Lett.} {\bfseries 48} (1982) 1063}.

\bibitem{Guralnik:2002ru}
Z.~Guralnik, R.~C. Helling, K.~Landsteiner, and E.~Lopez, ``{Perturbative instabilities on the noncommutative torus, Morita duality and twisted boundary conditions},'' \href{http://dx.doi.org/10.1088/1126-6708/2002/05/025}{{\em JHEP} {\bfseries 05} (2002) 025}, \href{http://arxiv.org/abs/hep-th/0204037}{{\ttfamily arXiv:hep-th/0204037}}.

\bibitem{Bietenholz:2006cz}
W.~Bietenholz, J.~Nishimura, Y.~Susaki, and J.~Volkholz, ``{A Non-perturbative study of 4-D U(1) non-commutative gauge theory: The Fate of one-loop instability},'' \href{http://dx.doi.org/10.1088/1126-6708/2006/10/042}{{\em JHEP} {\bfseries 10} (2006) 042}, \href{http://arxiv.org/abs/hep-th/0608072}{{\ttfamily arXiv:hep-th/0608072}}.

\bibitem{Teper:2006sp}
M.~Teper and H.~Vairinhos, ``{Symmetry breaking in twisted Eguchi\textendash{}Kawai models},'' \href{http://dx.doi.org/10.1016/j.physletb.2007.06.037}{{\em Phys. Lett. B} {\bfseries 652} (2007) 359--369}, \href{http://arxiv.org/abs/hep-th/0612097}{{\ttfamily arXiv:hep-th/0612097}}.

\bibitem{Azeyanagi:2007su}
T.~Azeyanagi, M.~Hanada, T.~Hirata, and T.~Ishikawa, ``{Phase structure of twisted Eguchi-Kawai model},'' \href{http://dx.doi.org/10.1088/1126-6708/2008/01/025}{{\em JHEP} {\bfseries 01} (2008) 025}, \href{http://arxiv.org/abs/0711.1925}{{\ttfamily arXiv:0711.1925 [hep-lat]}}.

\bibitem{Azeyanagi:2010ne}
T.~Azeyanagi, M.~Hanada, M.~Unsal, and R.~Yacoby, ``{Large-N reduction in QCD-like theories with massive adjoint fermions},'' \href{http://dx.doi.org/10.1103/PhysRevD.82.125013}{{\em Phys. Rev. D} {\bfseries 82} (2010) 125013}, \href{http://arxiv.org/abs/1006.0717}{{\ttfamily arXiv:1006.0717 [hep-th]}}.

\bibitem{Gonzalez-Arroyo:2014dua}
A.~Gonzalez-Arroyo and M.~Okawa, ``{Testing volume independence of SU(N) pure gauge theories at large N},'' \href{http://dx.doi.org/10.1007/JHEP12(2014)106}{{\em JHEP} {\bfseries 12} (2014) 106}, \href{http://arxiv.org/abs/1410.6405}{{\ttfamily arXiv:1410.6405 [hep-lat]}}.

\bibitem{Perez:2017jyq}
M.~Garc\'\i{}a~P\'erez, A.~Gonz\'alez-Arroyo, and M.~Okawa, ``{Perturbative contributions to Wilson loops in twisted lattice boxes and reduced models},'' \href{http://dx.doi.org/10.1007/JHEP10(2017)150}{{\em JHEP} {\bfseries 10} (2017) 150}, \href{http://arxiv.org/abs/1708.00841}{{\ttfamily arXiv:1708.00841 [hep-lat]}}.

\bibitem{Chamizo:2016msz}
F.~Chamizo and A.~Gonzalez-Arroyo, ``{Tachyonic instabilities in 2 + 1 dimensional Yang\textendash{}Mills theory and its connection to number theory},'' \href{http://dx.doi.org/10.1088/1751-8121/aa7346}{{\em J. Phys. A} {\bfseries 50} no.~26, (2017) 265401}, \href{http://arxiv.org/abs/1610.07972}{{\ttfamily arXiv:1610.07972 [hep-th]}}.

\bibitem{GarciaPerez:2018fkj}
M.~Garc\'\i{}a~P\'erez, A.~Gonz\'alez-Arroyo, M.~Koren, and M.~Okawa, ``{The spectrum of 2+1 dimensional Yang-Mills theory on a twisted spatial torus},'' \href{http://dx.doi.org/10.1007/JHEP07(2018)169}{{\em JHEP} {\bfseries 07} (2018) 169}, \href{http://arxiv.org/abs/1807.03481}{{\ttfamily arXiv:1807.03481 [hep-th]}}.

\bibitem{Bribian:2019ybc}
E.~I. Bribian and M.~Garcia~Perez, ``{The twisted gradient flow coupling at one loop},'' \href{http://dx.doi.org/10.1007/JHEP03(2019)200}{{\em JHEP} {\bfseries 03} (2019) 200}, \href{http://arxiv.org/abs/1903.08029}{{\ttfamily arXiv:1903.08029 [hep-lat]}}.

\bibitem{Gonzalez-Arroyo:2012euf}
A.~Gonzalez-Arroyo and M.~Okawa, ``{The string tension from smeared Wilson loops at large N},'' \href{http://dx.doi.org/10.1016/j.physletb.2012.12.027}{{\em Phys. Lett. B} {\bfseries 718} (2013) 1524--1528}, \href{http://arxiv.org/abs/1206.0049}{{\ttfamily arXiv:1206.0049 [hep-th]}}.

\bibitem{GarciaPerez:2014azn}
M.~Garc\'\i{}a~P\'erez, A.~Gonz\'alez-Arroyo, L.~Keegan, and M.~Okawa, ``{The $SU(\infty)$ twisted gradient flow running coupling},'' \href{http://dx.doi.org/10.1007/JHEP01(2015)038}{{\em JHEP} {\bfseries 01} (2015) 038}, \href{http://arxiv.org/abs/1412.0941}{{\ttfamily arXiv:1412.0941 [hep-lat]}}.

\bibitem{Gonzalez-Arroyo:2015bya}
A.~Gonz\'alez-Arroyo and M.~Okawa, ``{Large N meson masses from a matrix model},'' \href{http://dx.doi.org/10.1016/j.physletb.2016.02.001}{{\em Phys. Lett. B} {\bfseries 755} (2016) 132--137}, \href{http://arxiv.org/abs/1510.05428}{{\ttfamily arXiv:1510.05428 [hep-lat]}}.

\bibitem{Perez:2020vbn}
M.~Garc\'\i{}a~P\'erez, A.~Gonz\'alez-Arroyo, and M.~Okawa, ``{Meson spectrum in the large $N$ limit},'' \href{http://dx.doi.org/10.1007/JHEP04(2021)230}{{\em JHEP} {\bfseries 04} (2021) 230}, \href{http://arxiv.org/abs/2011.13061}{{\ttfamily arXiv:2011.13061 [hep-lat]}}.

\bibitem{Bonanno:2023ypf}
C.~Bonanno, P.~Butti, M.~Garc\'\i{}a~Per\'ez, A.~Gonz\'alez-Arroyo, K.-I. Ishikawa, and M.~Okawa, ``{The large-N limit of the chiral condensate from twisted reduced models},'' \href{http://dx.doi.org/10.1007/JHEP12(2023)034}{{\em JHEP} {\bfseries 12} (2023) 034}, \href{http://arxiv.org/abs/2309.15540}{{\ttfamily arXiv:2309.15540 [hep-lat]}}.

\bibitem{Bonanno:2024bqg}
C.~Bonanno, P.~Butti, M.~Garc\'\i{}a~P\'erez, A.~Gonz\'alez-Arroyo, K.-I. Ishikawa, and M.~Okawa, ``{Nonperturbative determination of the $\mathcal{N}=1$ supersymmetric Yang-Mills gluino condensate at large $N$},'' \href{http://dx.doi.org/10.1103/PhysRevD.110.074507}{{\em Phys. Rev. D} {\bfseries 110} no.~7, (2024) 074507}, \href{http://arxiv.org/abs/2406.08995}{{\ttfamily arXiv:2406.08995 [hep-th]}}.

\bibitem{Bonanno:2024onr}
C.~Bonanno, M.~Garc\'\i{}a~P\'erez, A.~Gonz\'alez-Arroyo, K.-I. Ishikawa, and M.~Okawa, ``{The mass of the gluino-glue bound state in large-$N$ $\mathcal{N} = 1$ Supersymmetric Yang-Mills theory},'' \href{http://dx.doi.org/10.1007/JHEP03(2025)174}{{\em JHEP} {\bfseries 03} (2025) 174}, \href{http://arxiv.org/abs/2412.02348}{{\ttfamily arXiv:2412.02348 [hep-lat]}}.

\bibitem{Davies:1999uw}
N.~M. Davies, T.~J. Hollowood, V.~V. Khoze, and M.~P. Mattis, ``{Gluino condensate and magnetic monopoles in supersymmetric gluodynamics},'' \href{http://dx.doi.org/10.1016/S0550-3213(99)00434-4}{{\em Nucl. Phys. B} {\bfseries 559} (1999) 123--142}, \href{http://arxiv.org/abs/hep-th/9905015}{{\ttfamily arXiv:hep-th/9905015}}.

\bibitem{Davies:2000nw}
N.~M. Davies, T.~J. Hollowood, and V.~V. Khoze, ``{Monopoles, affine algebras and the gluino condensate},'' \href{http://dx.doi.org/10.1063/1.1586477}{{\em J. Math. Phys.} {\bfseries 44} (2003) 3640--3656},
\href{http://arxiv.org/abs/hep-th/0006011}{{\ttfamily arXiv:hep-th/0006011 [hep-th]}}.

\bibitem{Unsal:2007vu}
M.~Unsal, ``{Abelian duality, confinement, and chiral symmetry breaking in QCD(adj)},'' \href{http://dx.doi.org/10.1103/PhysRevLett.100.032005}{{\em Phys. Rev. Lett.} {\bfseries 100} (2008) 032005},
\href{http://arxiv.org/abs/0708.1772}{{\ttfamily arXiv:0708.1772 [hep-th]}}.

\bibitem{Unsal:2007jx}
M.~Unsal, ``{Magnetic bion condensation: A New mechanism of confinement and mass gap in four dimensions},'' \href{http://dx.doi.org/10.1103/PhysRevD.80.065001}{{\em Phys. Rev.} {\bfseries D80} (2009) 065001},
\href{http://arxiv.org/abs/0709.3269}{{\ttfamily arXiv:0709.3269 [hep-th]}}.

\bibitem{Unsal:2008ch}
M.~Unsal and L.~G. Yaffe, ``{Center-stabilized Yang-Mills theory: Confinement and large N volume independence},'' \href{http://dx.doi.org/10.1103/PhysRevD.78.065035}{{\em Phys. Rev.} {\bfseries D78} (2008) 065035},
\href{http://arxiv.org/abs/0803.0344}{{\ttfamily arXiv:0803.0344 [hep-th]}}.

\bibitem{Shifman:2008ja}
M.~Shifman and M.~Unsal, ``{QCD-like Theories on R(3) x S(1): A Smooth Journey from Small to Large r(S(1)) with Double-Trace Deformations},'' \href{http://dx.doi.org/10.1103/PhysRevD.78.065004}{{\em Phys. Rev.} {\bfseries D78} (2008) 065004},
\href{http://arxiv.org/abs/0802.1232}{{\ttfamily arXiv:0802.1232 [hep-th]}}.

\bibitem{Poppitz:2008hr}
E.~Poppitz and M.~Unsal, ``{Index theorem for topological excitations on R**3 x S**1 and Chern-Simons theory},'' \href{http://dx.doi.org/10.1088/1126-6708/2009/03/027}{{\em JHEP} {\bfseries 03} (2009) 027},
\href{http://arxiv.org/abs/0812.2085}{{\ttfamily arXiv:0812.2085 [hep-th]}}.

\bibitem{Poppitz:2012sw}
E.~Poppitz, T.~Sch\"{a}fer, and M.~\"{U}nsal, ``{Continuity, Deconfinement, and (Super) Yang-Mills Theory},'' \href{http://dx.doi.org/10.1007/JHEP10(2012)115}{{\em JHEP} {\bfseries 10} (2012) 115},
\href{http://arxiv.org/abs/1205.0290}{{\ttfamily arXiv:1205.0290 [hep-th]}}.

\bibitem{Guvendik:2024yzh}
C.~Guvendik, ``{Towards Stabilization on Noncommutative Torus},'' \href{http://arxiv.org/abs/2411.12838}{{\ttfamily arXiv:2411.12838 [hep-th]}}.

\bibitem{tHooft:1981nnx}
G.~'t~Hooft, ``{Some Twisted Selfdual Solutions for the Yang-Mills Equations on a Hypertorus},'' \href{http://dx.doi.org/10.1007/BF01208900}{{\em Commun. Math. Phys.} {\bfseries 81} (1981) 267--275}.

\bibitem{Anber:2022qsz}
M.~M. Anber and E.~Poppitz, ``The gaugino condensate from asymmetric four-torus with twists,'' \href{http://dx.doi.org/10.1007/JHEP01(2023)118}{{\em JHEP} {\bfseries 01} (2023) 118}, \href{http://arxiv.org/abs/2210.13568}{{\ttfamily arxiv:2210.13568 [hep-th]}}.

\bibitem{Anber:2023sjn}
M.~M. Anber and E.~Poppitz, ``Multi-fractional instantons in {{SU}}({{N}}) {{Yang-Mills}} theory on the twisted {{T}}{$^{4}$},'' \href{http://dx.doi.org/10.1007/JHEP09(2023)095}{{\em JHEP} {\bfseries 09} (2023) 095}, \href{http://arxiv.org/abs/2307.04795}{{\ttfamily arxiv:2307.04795 [hep-th]}}.

\bibitem{Anber:2024mco}
M.~M. Anber and E.~Poppitz, ``{Higher-order gaugino condensates on a twisted $ {\mathbbm{T}}^4 $},'' \href{http://dx.doi.org/10.1007/JHEP02(2025)114}{{\em JHEP} {\bfseries 02} (2025) 114}, \href{http://arxiv.org/abs/2408.16058}{{\ttfamily arXiv:2408.16058 [hep-th]}}.

\bibitem{Anber:2025yub}
M.~M. Anber, A.~A. Cox, and E.~Poppitz, ``{On the moduli space of multi-fractional instantons on the twisted $\mathbb T^4$},'' \href{http://arxiv.org/abs/2504.06344}{{\ttfamily arXiv:2504.06344 [hep-th]}}.

\bibitem{Ford:2002pa}
C.~Ford and J.~M. Pawlowski, ``{Constituents of doubly periodic instantons},'' \href{http://dx.doi.org/10.1016/S0370-2693(02)02130-5}{{\em Phys. Lett. B} {\bfseries 540} (2002) 153--158}, \href{http://arxiv.org/abs/hep-th/0205116}{{\ttfamily arXiv:hep-th/0205116}}.

\bibitem{Ford:2003vi}
C.~Ford and J.~M. Pawlowski, ``{Doubly periodic instantons and their constituents},'' \href{http://dx.doi.org/10.1103/PhysRevD.69.065006}{{\em Phys. Rev. D} {\bfseries 69} (2004) 065006}, \href{http://arxiv.org/abs/hep-th/0302117}{{\ttfamily arXiv:hep-th/0302117}}.

\bibitem{Ford:2005sq}
C.~Ford and J.~M. Pawlowski, ``{Doubly periodic instanton zero modes},'' \href{http://dx.doi.org/10.1016/j.physletb.2005.08.094}{{\em Phys. Lett. B} {\bfseries 626} (2005) 139--146}, \href{http://arxiv.org/abs/hep-th/0505214}{{\ttfamily arXiv:hep-th/0505214}}.

\bibitem{GarciaPerez:1989gt}
M.~Garcia~Perez, A.~Gonzalez-Arroyo, and B.~Soderberg, ``{Minimum Action Solutions for SU(2) Gauge Theory on the Torus With Nonorthogonal Twist},'' \href{http://dx.doi.org/10.1016/0370-2693(90)90106-G}{{\em Phys. Lett. B} {\bfseries 235} (1990) 117--123}.

\bibitem{GarciaPerez:1992fj}
M.~Garcia~Perez and A.~{Gonzalez-Arroyo}, ``Numerical study of {{Yang-Mills}} classical solutions on the twisted torus,'' \href{http://dx.doi.org/10.1088/0305-4470/26/11/015}{{\em J. Phys. A} {\bfseries 26} no.~FTUAM-92-08, (1993) 2667--2678}, \href{http://arxiv.org/abs/hep-lat/9206016}{{\ttfamily arxiv:hep-lat/9206016}}.

\bibitem{Itou:2018wkm}
E.~Itou, ``Fractional instanton of the {{SU}}(3) gauge theory in weak coupling regime,'' \href{http://dx.doi.org/10.1007/JHEP05(2019)093}{{\em JHEP} {\bfseries 05} (2019) 093}, \href{http://arxiv.org/abs/1811.05708}{{\ttfamily arxiv:1811.05708 [hep-th]}}.

\bibitem{DasilvaGolan:2022jlm}
J.~Dasilva~Gol\'an and M.~Garc\'\i{}a~P\'erez, ``{SU(N) fractional instantons and the Fibonacci sequence},'' \href{http://dx.doi.org/10.1007/JHEP12(2022)109}{{\em JHEP} {\bfseries 12} (2022) 109}, \href{http://arxiv.org/abs/2208.07133}{{\ttfamily arXiv:2208.07133 [hep-th]}}.

\bibitem{Wandler:2024hsq}
F.~D. Wandler, ``{Numerical fractional instantons in SU(2): center vortices, monopoles, and a sharp transition between them},'' \href{http://arxiv.org/abs/2406.07636}{{\ttfamily arXiv:2406.07636 [hep-lat]}}.

\bibitem{Gonzalez-Arroyo:1998hjb}
A.~Gonzalez-Arroyo and A.~Montero, ``{Selfdual vortex - like configurations in SU(2) Yang-Mills theory},'' \href{http://dx.doi.org/10.1016/S0370-2693(98)01229-5}{{\em Phys. Lett. B} {\bfseries 442} (1998) 273--278}, \href{http://arxiv.org/abs/hep-th/9809037}{{\ttfamily arXiv:hep-th/9809037}}.

\bibitem{Montero:1999by}
A.~Montero, ``{Study of SU(3) vortex - like configurations with a new maximal center gauge fixing method},'' \href{http://dx.doi.org/10.1016/S0370-2693(99)01113-2}{{\em Phys. Lett. B} {\bfseries 467} (1999) 106--111}, \href{http://arxiv.org/abs/hep-lat/9906010}{{\ttfamily arXiv:hep-lat/9906010}}.

\bibitem{Montero:2000pb}
A.~Montero, ``{Vortex configurations in the large N limit},'' \href{http://dx.doi.org/10.1016/S0370-2693(00)00572-4}{{\em Phys. Lett. B} {\bfseries 483} (2000) 309--314}, \href{http://arxiv.org/abs/hep-lat/0004002}{{\ttfamily arXiv:hep-lat/0004002}}.

\bibitem{Lee:1997vp}
K.-M. Lee and P.~Yi, ``{Monopoles and instantons on partially compactified D-branes},'' \href{http://dx.doi.org/10.1103/PhysRevD.56.3711}{{\em Phys. Rev.} {\bfseries D56} (1997) 3711--3717},
\href{http://arxiv.org/abs/hep-th/9702107}{{\ttfamily arXiv:hep-th/9702107 [hep-th]}}.

\bibitem{Lee:1998bb}
K.-M. Lee and C.-h. Lu, ``{SU(2) calorons and magnetic monopoles},'' \href{http://dx.doi.org/10.1103/PhysRevD.58.025011}{{\em Phys. Rev.} {\bfseries D58} (1998) 025011},
\href{http://arxiv.org/abs/hep-th/9802108}{{\ttfamily arXiv:hep-th/9802108 [hep-th]}}.

\bibitem{Lee:1998vu}
K.-M. Lee, ``{Instantons and magnetic monopoles on R**3 x S**1 with arbitrary simple gauge groups},'' \href{http://dx.doi.org/10.1016/S0370-2693(98)00283-4}{{\em Phys. Lett.} {\bfseries B426} (1998) 323--328},
\href{http://arxiv.org/abs/hep-th/9802012}{{\ttfamily arXiv:hep-th/9802012 [hep-th]}}.

\bibitem{Kraan:1998kp}
T.~C. Kraan and P.~van Baal, ``{Exact T duality between calorons and Taub - NUT spaces},'' \href{http://dx.doi.org/10.1016/S0370-2693(98)00411-0}{{\em Phys. Lett.} {\bfseries B428} (1998) 268--276},
\href{http://arxiv.org/abs/hep-th/9802049}{{\ttfamily arXiv:hep-th/9802049 [hep-th]}}.

\bibitem{Kraan:1998pm}
T.~C. Kraan and P.~van Baal, ``{Periodic instantons with nontrivial holonomy},'' \href{http://dx.doi.org/10.1016/S0550-3213(98)00590-2}{{\em Nucl. Phys.} {\bfseries B533} (1998) 627--659},
\href{http://arxiv.org/abs/hep-th/9805168}{{\ttfamily arXiv:hep-th/9805168 [hep-th]}}.

\bibitem{Kraan:1998sn}
T.~C. Kraan and P.~van Baal, ``{Monopole constituents inside SU(n) calorons},'' \href{http://dx.doi.org/10.1016/S0370-2693(98)00799-0}{{\em Phys. Lett.} {\bfseries B435} (1998) 389--395},
\href{http://arxiv.org/abs/hep-th/9806034}{{\ttfamily arXiv:hep-th/9806034 [hep-th]}}.

\bibitem{Gross:1980br}
D.~J. Gross, R.~D. Pisarski, and L.~G. Yaffe, ``{QCD and Instantons at Finite Temperature},''
\href{http://dx.doi.org/10.1103/RevModPhys.53.43}{{\em Rev. Mod. Phys.} {\bfseries 53} (1981) 43}.

\bibitem{tHooft:1977nqb}
G.~'t~Hooft, ``On the phase transition towards permanent quark confinement,'' \href{http://dx.doi.org/10.1016/0550-3213(78)90153-0}{{\em Nucl.Phys.B} {\bfseries 138} (1978) 1--25}.

\bibitem{Cornwall:1979hz}
J.~M. Cornwall, ``Quark confinement and vortices in massive gauge invariant {{QCD}},'' \href{http://dx.doi.org/10.1016/0550-3213(79)90111-1}{{\em Nucl. Phys. B} {\bfseries 157} no.~UCLA/79/TEP/5, (1979) 392--412}.

\bibitem{Nielsen:1979xu}
H.~B. Nielsen and P.~Olesen, ``A quantum liquid model for the {{QCD}} vacuum: {{Gauge}} and rotational invariance of domained and quantized homogeneous color fields,'' \href{http://dx.doi.org/10.1016/0550-3213(79)90065-8}{{\em Nucl. Phys. B} {\bfseries 160} no.~NBI-HE-79-17, (1979) 380--396}.

\bibitem{Ambjorn:1980ms}
J.~Ambjorn and P.~Olesen, ``A color magnetic vortex condensate in {{QCD}},'' \href{http://dx.doi.org/10.1016/0550-3213(80)90150-9}{{\em Nucl. Phys. B} {\bfseries 170} no.~NBI-HE-80-14, (1980) 265--282}.

\bibitem{DelDebbio:1996lih}
L.~Del~Debbio, M.~Faber, J.~Greensite, and S.~Olejnik, ``Center dominance and {{Z}}(2) vortices in {{SU}}(2) lattice gauge theory,'' \href{http://dx.doi.org/10.1103/PhysRevD.55.2298}{{\em Phys. Rev. D} {\bfseries 55} no.~LBL-39424, LBNL-39424, SWAT-96-119, (1997) 2298--2306}, \href{http://arxiv.org/abs/hep-lat/9610005}{{\ttfamily arxiv:hep-lat/9610005}}.

\bibitem{Faber:1997rp}
M.~Faber, J.~Greensite, and S.~Olejnik, ``Casimir scaling from center vortices: {{Towards}} an understanding of the adjoint string tension,'' \href{http://dx.doi.org/10.1103/PhysRevD.57.2603}{{\em Phys. Rev. D} {\bfseries 57} (1998) 2603--2609}, \href{http://arxiv.org/abs/hep-lat/9710039}{{\ttfamily arxiv:hep-lat/9710039}}.

\bibitem{DelDebbio:1998luz}
L.~Del~Debbio, M.~Faber, J.~Giedt, J.~Greensite, and S.~Olejnik, ``Detection of center vortices in the lattice {{Yang-Mills}} vacuum,'' \href{http://dx.doi.org/10.1103/PhysRevD.58.094501}{{\em Phys. Rev. D} {\bfseries 58} (1998) 094501}, \href{http://arxiv.org/abs/hep-lat/9801027}{{\ttfamily arxiv:hep-lat/9801027}}.

\bibitem{Langfeld:1998cz}
K.~Langfeld, O.~Tennert, M.~Engelhardt, and H.~Reinhardt, ``Center vortices of {{Yang-Mills}} theory at finite temperatures,'' \href{http://dx.doi.org/10.1016/S0370-2693(99)00252-X}{{\em Phys. Lett. B} {\bfseries 452} no.~UNITU-THEP-5-98, (1999) 301}, \href{http://arxiv.org/abs/hep-lat/9805002}{{\ttfamily arxiv:hep-lat/9805002}}.

\bibitem{Kovacs:1998xm}
T.~G. Kovacs and E.~T. Tomboulis, ``Vortices and confinement at weak coupling,'' \href{http://dx.doi.org/10.1103/PhysRevD.57.4054}{{\em Phys. Rev. D} {\bfseries 57} no.~UCLA-97-TEP-22, COLO-HEP-392, (1998) 4054--4062}, \href{http://arxiv.org/abs/hep-lat/9711009}{{\ttfamily arxiv:hep-lat/9711009}}.

\bibitem{Engelhardt:1999fd}
M.~Engelhardt, K.~Langfeld, H.~Reinhardt, and O.~Tennert, ``Deconfinement in {{SU}}(2) {{Yang-Mills}} theory as a center vortex percolation transition,'' \href{http://dx.doi.org/10.1103/PhysRevD.61.054504}{{\em Phys. Rev. D} {\bfseries 61} (2000) 054504}, \href{http://arxiv.org/abs/hep-lat/9904004}{{\ttfamily arxiv:hep-lat/9904004}}.

\bibitem{deForcrand:1999our}
P.~{de Forcrand} and M.~D'Elia, ``On the relevance of center vortices to {{QCD}},'' \href{http://dx.doi.org/10.1103/PhysRevLett.82.4582}{{\em Phys. Rev. Lett.} {\bfseries 82} (1999) 4582--4585}, \href{http://arxiv.org/abs/hep-lat/9901020}{{\ttfamily arxiv:hep-lat/9901020}}.

\bibitem{vanBaal:1982ag}
P.~van Baal, ``{Some Results for SU(N) Gauge Fields on the Hypertorus},''
\href{http://dx.doi.org/10.1007/BF01403503}{{\em Commun. Math. Phys.} {\bfseries 85} (1982) 529}.

\bibitem{tHooft:1974kcl}
G.~'t~Hooft, ``{Magnetic Monopoles in Unified Gauge Theories},''
\href{http://dx.doi.org/10.1016/0550-3213(74)90486-6}{{\em Nucl. Phys.} {\bfseries B79} (1974) 276--284}.

\bibitem{Polyakov:1974ek}
A.~M. Polyakov, ``{Particle Spectrum in the Quantum Field Theory},''
{\em JETP Lett.} {\bfseries 20} (1974) 194--195.

\bibitem{Anber:2015wha}
M.~M. Anber and E.~Poppitz, ``{On the global structure of deformed Yang-Mills theory and QCD(adj) on $ {\mathrm{\mathbb{R}}}^3\times {\mathbb{S}}^1 $},'' \href{http://dx.doi.org/10.1007/JHEP10(2015)051}{{\em JHEP} {\bfseries 10} (2015) 051}, \href{http://arxiv.org/abs/1508.00910}{{\ttfamily arXiv:1508.00910 [hep-th]}}.

\bibitem{DelDebbio:1997ke}
L.~Del~Debbio, M.~Faber, J.~Greensite, and S.~Olejnik, ``{Center dominance, center vortices, and confinement},'' in {\em {NATO Advanced Research Workshop on Theoretical Physics: New Developments in Quantum Field Theory}}, pp.~47--64.
\newblock 6, 1997.
\newblock \href{http://arxiv.org/abs/hep-lat/9708023}{{\ttfamily arXiv:hep-lat/9708023}}.

\bibitem{Ambjorn:1999ym}
J.~Ambjorn, J.~Giedt, and J.~Greensite, ``Vortex structure versus monopole dominance in {{Abelian}} projected gauge theory,'' \href{http://dx.doi.org/10.1088/1126-6708/2000/02/033}{{\em JHEP} {\bfseries 02} (2000) 033}, \href{http://arxiv.org/abs/hep-lat/9907021}{{\ttfamily arxiv:hep-lat/9907021}}.

\bibitem{deForcrand:2000pg}
P.~de~Forcrand and M.~Pepe, ``{Center vortices and monopoles without lattice Gribov copies},'' \href{http://dx.doi.org/10.1016/S0550-3213(01)00009-8}{{\em Nucl. Phys. B} {\bfseries 598} (2001) 557--577}, \href{http://arxiv.org/abs/hep-lat/0008016}{{\ttfamily arXiv:hep-lat/0008016}}.

\bibitem{Engelhardt:1999xw}
M.~Engelhardt and H.~Reinhardt, ``{Center projection vortices in continuum Yang-Mills theory},'' \href{http://dx.doi.org/10.1016/S0550-3213(99)00727-0}{{\em Nucl. Phys. B} {\bfseries 567} (2000) 249}, \href{http://arxiv.org/abs/hep-th/9907139}{{\ttfamily arXiv:hep-th/9907139}}.

\bibitem{Reinhardt:2001kf}
H.~Reinhardt, ``{Topology of center vortices},'' \href{http://dx.doi.org/10.1016/S0550-3213(02)00130-X}{{\em Nucl. Phys. B} {\bfseries 628} (2002) 133--166}, \href{http://arxiv.org/abs/hep-th/0112215}{{\ttfamily arXiv:hep-th/0112215}}.

\bibitem{Cornwall:1999xw}
J.~M. Cornwall, ``{Center vortices, nexuses, and fractional topological charge},'' \href{http://dx.doi.org/10.1103/PhysRevD.61.085012}{{\em Phys. Rev. D} {\bfseries 61} (2000) 085012}, \href{http://arxiv.org/abs/hep-th/9911125}{{\ttfamily arXiv:hep-th/9911125}}.

\bibitem{Luscher:1981zq}
M.~Luscher, ``{Topology of Lattice Gauge Fields},'' \href{http://dx.doi.org/10.1007/BF02029132}{{\em Commun. Math. Phys.} {\bfseries 85} (1982) 39}.

\bibitem{Abe:2023ncy}
M.~Abe, O.~Morikawa, S.~Onoda, H.~Suzuki, and Y.~Tanizaki, ``{Topology of SU(N) lattice gauge theories coupled with \ensuremath{\mathbb{Z}}$_{N}$ 2-form gauge fields},'' \href{http://dx.doi.org/10.1007/JHEP08(2023)118}{{\em JHEP} {\bfseries 08} (2023) 118}, \href{http://arxiv.org/abs/2303.10977}{{\ttfamily arXiv:2303.10977 [hep-lat]}}.

\bibitem{Tanizaki:2017bam}
Y.~Tanizaki and Y.~Kikuchi, ``{Vacuum structure of bifundamental gauge theories at finite topological angles},'' \href{http://dx.doi.org/10.1007/JHEP06(2017)102}{{\em JHEP} {\bfseries 06} (2017) 102},
\href{http://arxiv.org/abs/1705.01949}{{\ttfamily arXiv:1705.01949 [hep-th]}}.

\bibitem{Kikuchi:2017pcp}
Y.~Kikuchi and Y.~Tanizaki, ``{Global inconsistency, 't~Hooft anomaly, and level crossing in quantum mechanics},'' \href{http://dx.doi.org/10.1093/ptep/ptx148}{{\em Prog. Theor. Exp. Phys.} {\bfseries 2017} (2017) 113B05},
\href{http://arxiv.org/abs/1708.01962}{{\ttfamily arXiv:1708.01962 [hep-th]}}.

\bibitem{Tanizaki:2018xto}
Y.~Tanizaki and T.~Sulejmanpasic, ``{Anomaly and global inconsistency matching: $\theta$-angles, $SU(3)/U(1)^2$ nonlinear sigma model, $SU(3)$ chains and its generalizations},'' \href{http://dx.doi.org/10.1103/PhysRevB.98.115126}{{\em Phys. Rev.} {\bfseries B98} no.~11, (2018) 115126},
\href{http://arxiv.org/abs/1805.11423}{{\ttfamily arXiv:1805.11423 [cond-mat.str-el]}}.

\bibitem{Cordova:2019jnf}
C.~Cordova, D.~S. Freed, H.~T. Lam, and N.~Seiberg, ``{Anomalies in the Space of Coupling Constants and Their Dynamical Applications I},'' \href{http://dx.doi.org/10.21468/SciPostPhys.8.1.001}{{\em SciPost Phys.} {\bfseries 8} no.~1, (2020) 001}, \href{http://arxiv.org/abs/1905.09315}{{\ttfamily arXiv:1905.09315 [hep-th]}}.

\bibitem{Cordova:2019uob}
C.~Cordova, D.~S. Freed, H.~T. Lam, and N.~Seiberg, ``{Anomalies in the Space of Coupling Constants and Their Dynamical Applications II},'' \href{http://dx.doi.org/10.21468/SciPostPhys.8.1.002}{{\em SciPost Phys.} {\bfseries 8} no.~1, (2020) 002}, \href{http://arxiv.org/abs/1905.13361}{{\ttfamily arXiv:1905.13361 [hep-th]}}.

\bibitem{Nguyen:2023fun}
M.~Nguyen, Y.~Tanizaki, and M.~\"Unsal, ``{Study of gapped phases of 4d gauge theories using temporal gauging of the~$\mathbb{Z}_N$ 1-form symmetry},'' \href{http://dx.doi.org/10.1007/JHEP08(2023)013}{{\em JHEP} {\bfseries 08} (2023) 013}, \href{http://arxiv.org/abs/2306.02485}{{\ttfamily arXiv:2306.02485 [hep-th]}}.

\bibitem{Kapustin:2013qsa}
A.~Kapustin and R.~Thorngren, ``{Topological Field Theory on a Lattice, Discrete Theta-Angles and Confinement},'' \href{http://dx.doi.org/10.4310/ATMP.2014.v18.n5.a4}{{\em Adv. Theor. Math. Phys.} {\bfseries 18} no.~5, (2014) 1233--1247}, \href{http://arxiv.org/abs/1308.2926}{{\ttfamily arXiv:1308.2926 [hep-th]}}.

\bibitem{Gukov:2013zka}
S.~Gukov and A.~Kapustin, ``{Topological Quantum Field Theory, Nonlocal Operators, and Gapped Phases of Gauge Theories},'' \href{http://arxiv.org/abs/1307.4793}{{\ttfamily arXiv:1307.4793 [hep-th]}}.

\bibitem{Barkeshli:2014cna}
M.~Barkeshli, P.~Bonderson, M.~Cheng, and Z.~Wang, ``{Symmetry Fractionalization, Defects, and Gauging of Topological Phases},'' \href{http://dx.doi.org/10.1103/PhysRevB.100.115147}{{\em Phys. Rev. B} {\bfseries 100} no.~11, (2019) 115147}, \href{http://arxiv.org/abs/1410.4540}{{\ttfamily arXiv:1410.4540 [cond-mat.str-el]}}.

\bibitem{Delmastro:2022pfo}
D.~G. Delmastro, J.~Gomis, P.-S. Hsin, and Z.~Komargodski, ``{Anomalies and symmetry fractionalization},'' \href{http://dx.doi.org/10.21468/SciPostPhys.15.3.079}{{\em SciPost Phys.} {\bfseries 15} no.~3, (2023) 079}, \href{http://arxiv.org/abs/2206.15118}{{\ttfamily arXiv:2206.15118 [hep-th]}}.

\bibitem{Poppitz:2011wy}
E.~Poppitz and M.~Unsal, ``{Seiberg-Witten and 'Polyakov-like' magnetic bion confinements are continuously connected},'' \href{http://dx.doi.org/10.1007/JHEP07(2011)082}{{\em JHEP} {\bfseries 07} (2011) 082},
\href{http://arxiv.org/abs/1105.3969}{{\ttfamily arXiv:1105.3969 [hep-th]}}.

\bibitem{Douglas:1995nw}
M.~R. Douglas and S.~H. Shenker, ``{Dynamics of SU(N) supersymmetric gauge theory},'' \href{http://dx.doi.org/10.1016/0550-3213(95)00258-T}{{\em Nucl. Phys. B} {\bfseries 447} (1995) 271--296}, \href{http://arxiv.org/abs/hep-th/9503163}{{\ttfamily arXiv:hep-th/9503163}}.

\bibitem{vanBaal:1983eq}
P.~van Baal, ``{Surviving Extrema for the Action on the Twisted SU(infinity) One Point Lattice},'' \href{http://dx.doi.org/10.1007/BF01206312}{{\em Commun. Math. Phys.} {\bfseries 92} (1983) 1}.

\end{thebibliography}\endgroup

\end{document}